\documentclass{aa} 

\usepackage{palatino,indentfirst}
\usepackage{float}
\usepackage[raggedright]{sidecap}
\usepackage{multicol,multirow,dcolumn,colortbl,pdflscape}
\usepackage{graphicx}
\usepackage{pdfpages}
\includepdfset{pages=-,link,pagecommand={\thispagestyle{fancy}}}
\usepackage[svgnames]{xcolor}
\usepackage{amsmath,amssymb}
\usepackage{threeparttable}
\usepackage{graphicx}
\usepackage[pdftex]{hyperref} % doit toujours tre charg en dernier (a part geometry): redfinit certaines commandes
\hypersetup{colorlinks=true,urlcolor=blue}
\usepackage{geometry}
\def\sun{\ensuremath\odot}
\usepackage{txfonts}
\begin{document}

 \title{The impact of convective criteria on the properties of massive stars} 
 \titlerunning{The impact of convective criteria on the properties of massive stars}
 \authorrunning{Sibony et al.}

 \author{Y. Sibony \and C. Georgy \and S. Ekström \and G. Meynet}

 \institute{Observatoire de Genève,
              Chemin Pegasi 51, 1290 Versoix, Switzerland\\\email{yves.sibony@unige.ch}}

\date{}

% 5 {} token are mandatory
 \abstract
 % context heading (optional)
 {Libraries of stellar models computed with either the Ledoux or the Schwarzschild criterion to determine the sizes of convective regions are available in the literature. It is still not clear, however, which of these two criteria should be used, although many works have been devoted to that question in the past.}
 {In the framework of the evolution of single rotating stars, we study the differences between models computed with Ledoux and Schwarzschild criteria on the internal structure, evolutionary track in the Hertzsprung-Russell diagram  (HRD), lifetimes, evolution of the surface abundances and velocities, and masses of the He and CO cores. We investigate the consequences on the nature of the supernova (SN) progenitors and the type of SN events, as well as on the stellar yields of light elements. We also study the impact on the outputs of population synthesis models.}
 {Models with initial masses between 7 and 120\,M$_\odot$ at solar metallicity ($Z$=0.014) and with an initial rotation equal to 0 or 0.4 times the critical velocity at the zero-age main sequence were computed with either the Schwarzschild or the Ledoux criterion until the end of the C-burning phase.}
 {Models with initial masses between 15 and 32 M$_\odot$ computed with the Schwarzschild criterion show larger intermediate convective zones attached to the H-burning shell than models computed with the Ledoux criterion. Their CO cores and outer convective zones in the red supergiant (RSG) phase are also smaller. This impacts many outputs of stars during the core He-burning phase. Schwarzschild models have smaller CO cores and outer convective zones in the RSG phase, and their blue-to-red supergiant ratio is much higher than for Ledoux models. They also produce longer crossings of the Hertzsprung gap and favour blue loops. The upper luminosity of RSGs is little affected by the change in the convective criterion. The maximum luminosity of RSG progenitors for type II-P SN events is lowered from 5.2 to 4.95 when the Ledoux criterion is used instead of the Schwarzschild criterion in non-rotating models. The Schwarzschild criterion predicts longer-lasting, less nitrogen-enriched, and faster-rotating Cepheids. Rotational mixing tends to decrease the differences between Schwarzschild and Ledoux models.}
 {The results of this paper can be used as first guidelines to set up observational programs that may help to distinguish between these two model families.}


 \maketitle
%-------------------------------------------------------------------

\section{Introduction}

Numerous difficulties arise when stars are modeled. The treatment of convection has long been one of these difficulties. The classical way of dealing with convection in one-dimensional (1D) stellar evolution codes is the following: 1) find the boundaries of the convective zone by applying an instability criterion, 2) empirically increase the size of (some of) the convective zones (a process often called ``overshooting''), and 3) compute a thermal gradient to be applied inside the convective zone. For the first point, two criteria are widely used in the literature to determine the stability of a thermally stratified medium with the gravity as the restoring force: the so-called Schwarzschild criterion, which writes (for stability against convection)
\begin{equation}
\nabla_\text{rad} < \nabla_\text{ad},\label{Schwarzschild}
\end{equation}
where $\nabla_\text{rad} = \left(\frac{\text{d}\ln T}{\text{d}\ln P}\right)_\text{rad}$ is the radiative thermal gradient, and $\nabla_\text{ad} = \left(\frac{\text{d}\ln T}{\text{d}\ln P}\right)_\text{ad}$ is the adiabatic thermal gradient; and the Ledoux criterion, which writes (also for stability)
\begin{equation}
\nabla_\text{rad} < \nabla_\text{ad} + \frac{\varphi}{\delta}\nabla_\mu,\label{Ledoux}
\end{equation}
where $\varphi = \left(\frac{\partial\ln\rho}{\partial\ln\mu}\right)_{T,P}$, $\delta = -\left(\frac{\partial\ln\rho}{\partial\ln T}\right)_{T,\mu}$, and $\nabla_\mu = \left(\frac{\text{d}\ln\mu}{\text{d}\ln P}\right)$.
It is possible that some layers inside the star are Ledoux-stable but Schwarzschild-unstable. In case of a thermally dissipative medium, however, \citet{Kato1966a} showed that Eq.~(\ref{Ledoux}) is reduced to Eq.~(\ref{Schwarzschild}) due to the onset of oscillatory convection. For this reason, the Schwarzschild criterion is most of the time preferred to the Ledoux criterion in stellar evolution computations. Another option is to apply a partial mixing of the layers encompassed by the two criteria. This is called ``semiconvection'' \citep[see][]{Langer1983a}. In this framework, models computed with the Ledoux criterion correspond to a totally inefficient semiconvection, while models computed with the Schwarzschild criterion correspond to an infinitely efficient semiconvection.

Point 2 was introduced in stellar evolution codes as a necessity to reproduce some observational features, such as the width of the main sequence (MS) of open clusters \citep[see e.g.][]{Maeder1981c,MM1989,Martinet2021}. Different implementations for this additional mixing (co)exist in stellar evolution codes: penetrative overshoot \citep[e.g.][]{Maeder1981c,Zahn1991a}, diffusive overshoot \citep[e.g.][]{Freytag1996a,Herwig2006a}, entrainment \citep{Meakin2007a,Cristini2017a,Scott2021a}, or more recently deduced from 2D fully compressible time-implicit simulations \citep{Baraffe2023}. A thorough discussion of these various implementations and their link to hydrodynamics simulation can be found in \citet{Viallet2015a}.

Finally, point 3 requires the computation of a thermal gradient to be applied inside the convective regions. For deep convection, the timescale for radiative exchanges between the convective cells and the environment are longer than the convective instability timescale, and the process is very close to being adiabatic, thus the adiabatic gradient $\nabla_\text{ad}$ can be applied. In convective regions closer to the surface, more complex theories need to be used. A usual choice in this case is the mixing-length theory \citep{Boehm-Vitense1958a}.

To obtain a complete picture of convection in stellar interiors, 3D hydrodynamics simulations are required. Convection in stars is very turbulent: the estimated Reynolds number is as high as $10^9$ \citep{Arnett2016a}. The mass flux is therefore highly asymmetric, requiring a multi-dimensional treatment. Efforts to do this have been made during the past decade for various physical environments \citep[e.g.][]{Meakin2007a,Cristini2017a,Woodward2015a,Jones2017a,Horst2021a}. However, all of these simulations were limited to a small region and/or a short time compared to the size and lifetime of a star. These results can nevertheless be used to build new algorithms that can obtain a better agreement between the 1D stellar evolution codes and the 3D convection simulations. Some attempts have been made \citep[e.g.][]{Scott2021a}, but these procedures are not yet fully mature and still need to be improved for routine use in stellar evolution calculations.

It is not clearly established which of the criteria for convection should be used in stellar evolution computations. Moreover, when the Ledoux criterion is used, the efficiency of semiconvection is not known either. Since the pioneering works of \citet{Stothers1973a,Stothers1975a,Stothers1976a}, very little progress has been made. \citet{Lawlor2015} simulated and compared the evolution and supernova (SN) light curves of Population III stars computed with either criterion and found Schwarzschild stars to be much larger and cooler than their Ledoux counterparts. \citet{Chun2018} calibrated the mixing-length parameter with red supergiants (RSGs) of different metallicities and predicted a decrease in the hydrogen content of stellar envelopes with increasing metallicity when using the Schwarzschild criterion, but not the Ledoux criterion. \citet{Schootemeijer2019} computed grids of Ledoux models at the metallicity of the Small Magellanic Cloud (SMC; $Z=0.002$) and found that more efficient semiconvection increases the ratio of blue to red supergiants. \citet{Kaiser2020} computed non-rotating models of 15, 20, and 25\,M$_\odot$ at solar metallicity to study the relative impact of various convective parameters. They found the strength of convective boundary mixing to be more effective than the choice of criterion for convection, except for determining the initial location of intermediate convective zones. \citet{Anders2022} performed a 3D hydrodynamical simulation of a convective zone adjacent to a semiconvective region. They reported that both criteria gave the same results when the evolutionary timescale exceeds the convective-overturn timescale (e.g. during the main sequence), but that differences can emerge when the evolution occurs on a rapid timescale (e.g. the Hertzsprung gap is crossed). We pursue our efforts and to try to determine which criterion reproduces the observed features of massive stars better \citep{Saio2013a,Georgy2014,Georgy2021}. To do this, we provide the first detailed comparison of Schwarzschild and Ledoux criterion models over a wide mass range at solar metallicity, including the effects of rotation.

In Sect.~\ref{Ingredients}, we present the main physical ingredients of the stellar models. We discuss the impact of changing the convective criterion on the cores and convective structures of stars in Sect.~\ref{internal_structure}. Section~\ref{Evolution} presents the impact on various aspects of post-MS stellar evolution. We compare the properties of stars at the end of their evolution in Sect.~\ref{final_properties}. We present models for population synthesis in Sect.~\ref{Populations}. We compare our results to previous works on the subject and to an observed population of evolved massive stars in Sect.~\ref{comparisons}. The discussion and conclusions are given in Sect.~\ref{Conclusion}.

\section{Ingredients of the models}
\label{Ingredients}

The models computed with the Schwarzschild criterion were presented in \citet{Ekstrom2012}. The convective regions are determined starting from the centre of the star and going outwards. In each shell of the model, the adiabatic and the radiative gradient are computed and compared, which determines whether the shell is to be radiative or convective. During central H- and He-burning, the size of the core is artificially increased by a length corresponding to 10\% of the local pressure scale height at the edge of the formal core (overshooting). This is not done for subsequent burning phases or for convective shells. In the internal convective zones, the thermal gradient is imposed to be adiabatic, and the chemical species are instantaneously mixed (so that the chemical composition of the convective zones is always homogeneous). In the envelope, the classical mixing-length theory \citep{Boehm-Vitense1958a} is used to obtain the thermal gradient, with a mixing-length parameter $\alpha = 1.6$. For models more massive than 40$\,M_\sun$, the mixing-length parameter is set to $1$, and the mixing-length is computed according to the density scale height instead of the pressure scale height. Moreover, the turbulent pressure is accounted for by adding an acoustic flux term in the mixing-length formulation \citep[see][]{Maeder1987c}.

The Ledoux models are new models that are computed with the same physical ingredients as the Schwarzschild models, except that the $\mu$-gradient is computed and accounted for when a shell is to be determined as radiative or convective. All the other ingredients are kept the same. In particular, the treatment of overshooting and of mixing is the same, and the mass-loss prescriptions are the same as well.

Rotation is also treated in the same way in both sets of models. The full advection-diffusion equation for the transport of angular momentum is solved during the main sequence \citep[see e.g.][for more details]{Eggenberger2008a}. The horizontal diffusion coefficient that intervenes in this formalism was proposed by \citet{Zahn1992a}, and the diffusion coefficient associated with the shear mixing was proposed by \citet{Maeder1997a}. All the rotating models shown in this paper have an initial rotation rate $\upsilon_\text{ini}/\upsilon_\text{crit} = 0.4$, where the critical velocity $\upsilon_\text{crit}$ is computed as in \citet[see their definition of $\upsilon_\text{crit,1}$]{Georgy2011a}.

Mass-loss is applied to all of our models according to different recipes, which are summarised here \citep[see also][]{Ekstrom2012}. For non-Wolf-Rayet (WR) stars hotter than $\log(T_\text{eff}) = 4.0$, the mass-loss rates by \citet{Vink2000a,Vink2001a} are applied. In this case, a correction factor accounting for the effects of rotation on the mass loss is also applied \citep[see e.g.][]{Georgy2011a}. For models up to 9$\,M_\sun$ and when $\log(T_\text{eff}) < 3.8$, the \citet{Reimers1977a} rates are used, with the parameter $\eta = 0.6$. For more massive models, the mass-loss rate is a linear fit of observational data from \citet{Sylvester1998a} and \citet[see also \citealt{Crowther2001a}]{vanLoon1999a}. When the surface abundance of H drops below $X_s < 0.3$ and the effective temperature is above $T_\text{eff} > 4.0$, we switch to WR-kind mass-loss rates \citep{Nugis2000a,Grafener2008a}, unless they are lower than the \citet{Vink2000a} rates. Finally, we apply the \citet{deJager1988a} rates when the above recipes are not applicable.

For massive RSGss, some layers in the external layers, very close to the surface, reach luminosities far above the Eddington luminosity. To facilitate the computation in this case, we therefore increase the mass-loss rate as described above by a factor of $3$ when the over-Eddington luminosity is overcome by a factor of $5$ for models more massive than 20\,M$_\sun$.

\begin{table}[!]
\caption{Properties of the cores and convective zones at different evolutionary stages, and the mass at the final stage.}
\centering
\scalebox{.65}{\begin{tabular}{|ccc|cc|c|ccc|c|}
\hline\hline
\multicolumn{3}{|c|}{} & \multicolumn{2}{c|}{Core properties} & \multicolumn{1}{c|}{ICZ} & \multicolumn{3}{c|}{OCZ} & M$_\text{fin}$\\ 
Crit. & M$_\text{ini}$ & ${\frac{\upsilon_\text{ini}}{\upsilon_\text{crit}}}$ & He & CO & & pre-He-b & He-b & post He-b &\\ 
& $M_{\sun}$ && \multicolumn{2}{c|}{ $M_\sun$} & $M_{\sun}$ & \multicolumn{3}{c|}{ $M_\sun$} & $M_\sun$\\ 
\hline
Led. & $120$ & $0.4$ & $55.6$ & $25.3$ & -- & -- & -- & -- & $30.8$\\ 
Sch. & $120$ & $0.4$ & $33.7$ & $14.9$ & -- & -- & -- & -- & $19.0$\\ 
Led. & $120$ & -- & $54.4$ & $23.5$ & $50.8-58.8$ & -- & -- & -- & $28.6$\\ 
Sch. & $120$ & -- & $55.2$ & $25.5$ & $50.8-59.8$ & -- & -- & -- & $30.9$\\ \hline
Led. & $85$ & $0.4$ & $42.7$ & $18.1$ & -- & -- & -- & -- & $22.7$\\ 
Sch. & $85$ & $0.4$ & $42.8$ & $21.4$ & -- & -- & -- & -- & $26.4$\\ 
Led. & $85$ & -- & $37.0$ & $14.8$ & $34.9-44.6$ & -- & -- & -- & $19.0$\\ 
Sch. & $85$ & -- & $37.4$ & $14.7$ & $35.0-44.0$ & -- & -- & -- & $18.6$\\ \hline
Led. & $60$ & $0.4$ & $30.6$ & $12.9$ & $27.4-34.4$ & -- & -- & -- & $16.8$\\ 
Sch. & $60$ & $0.4$ & $31.4$ & $14.0$ & $27.6-34.1$ & -- & - & -- & $18.0$\\ 
Led. & $60$ & -- & $24.5$ & $10.4$ & $25.2-33.6$ & -- & -- & -- & $13.9$\\ 
Sch. & $60$ & -- & $24.9$ & $9.14$ & $23.2-33.5$ & -- & -- & -- & $12.5$\\ \hline
Led. & $40$ & $0.4$ & $18.5$ & $11.6$ & $17.0-27.7$ & $30.7$ & $30.1$ & -- & $15.4$\\ 
Sch. & $40$ & $0.4$ & $18.7$ & $8.98$ & $18.0-29.6$ & $30.4$ & $29.3$ & -- & $12.3$\\ 
Led. & $40$ & -- & $14.6$ & $9.41$ & $14.0-22.2$ & -- & $23.8$ & -- & $12.9$\\ 
Sch. & $40$ & -- & $14.8$ & $10.4$ & $13.7-25.3$ & -- & $31.8$ & -- & $13.7$\\ \hline
Led. & $32$ & $0.4$ & $13.5$ & $10.5$ & $13.3-22.9$ & -- & -- & -- & $14.1$\\ 
Sch. & $32$ & $0.4$ & $13.5$ & $7.03$ & $12.4-25.8$ & -- & -- & -- & $10.1$\\ 
Led. & $32$ & -- & $10.8$ & $8.29$ & $17.2-21.6$ & -- & $16.0$ & -- & $11.3$\\ 
Sch. & $32$ & -- & $11.0$ & $7.70$ & $10.6-23.3$ & -- & $25.5$ & -- & $10.9$\\ \hline
Led. & $25$ & $0.4$ & $9.1$ & $6.91$ & $7.0-16.8$ & -- & $10.7$ & -- & $10.4$\\ 
Sch. & $25$ & $0.4$ & $9.1$ & $6.53$ & $8.6-17.5$ & -- & $19.7$ & -- & $9.7$\\ 
Led. & $25$ & -- & $7.6$ & $6.23$ & $12.0-15.6$ & -- & $9.8$ & -- & $9.3$\\ 
Sch. & $25$ & -- & $7.7$ & $5.29$ & $7.5-17.2$ & -- & $18.7$ & -- & $8.3$\\ \hline
Led. & $20$ & $0.4$ & $6.2$ & $4.44$ & $7.4-12.4$ & -- & $7.2$ & $7.6$ & $7.3$\\ 
Sch. & $20$ & $0.4$ & $6.3$ & $4.36$ & $5.6-12.5$ & -- & $14.6$ & -- & $7.2$\\ 
Led. & $20$ & -- & $5.3$ & $4.18$ & $6.1-12.3$ & -- & $8.2$ & -- & $7.0$\\ 
Sch. & $20$ & -- & $5.4$ & $3.65$ & $5.5-12.6$ & -- & $13.4$ & $6.4$ & $8.6$\\ \hline
Led. & $15$ & $0.4$ & $3.8$ & $2.66$ & -- & $5.8$ & $5.1$ & $5.2$ & $12.4$\\ 
Sch. & $15$ & $0.4$ & $3.8$ & $2.68$ & $3.9-7.5$ & -- & $5.9$ & $5.2$ & $11.1$\\ 
Led. & $15$ & -- & $3.3$ & $2.28$ & $6.2-8.4$ & $5.0$ & $4.8$ & $4.6$ & $13.0$\\ 
Sch. & $15$ & -- & $3.3$ & $2.11$ & $3.5-7.6$ & -- & $5.8$ & $4.3$ & $13.2$\\ \hline
Led. & $12$ & $0.4$ & $2.5$ & $1.69$ & -- & $3.6$ & $3.6$ & $3.8$ & $11.1$\\ 
Sch. & $12$ & $0.4$ & $2.5$ & $1.82$ & $2.8-5.1$ & $4.9$ & $4.9$ & $3.9$ & $10.2$\\ 
Led. & $12$ & -- & $2.3$ & $1.43$ & $4.4-5.5$ & $3.9$ & $3.5$ & $3.3$ & $11.1$\\ 
Sch. & $12$ & -- & $2.3$ & $1.26$ & $2.6-5.0$ & $2.6$ & $3.2$ & $3.0$ & $11.3$\\ \hline
Led. & $9$ & $0.4$ & $1.5$ & $0.93$ & -- & $2.4$ & $2.3$ & $2.5$ & $8.7$\\ 
Sch. & $9$ & $0.4$ & $1.5$ & $1.18$ & -- & $2.4$ & $2.3$ & $3.1$ & $8.5$\\ 
Led. & $9$ & -- & $1.4$ & $0.82$ & -- & $2.4$ & $2.4$ & $2.3$ & $8.6$\\ 
Sch. & $9$ & -- & $1.4$ & $0.83$ & -- & $2.3$ & $2.2$ & $1.2$ & $8.8$\\ \hline
Led. & $7$ & $0.4$ & $1.0$ & $0.60$ & -- & $1.7$ & $1.7$ & $1.2$ & $6.9$\\ 
Sch. & $7$ & $0.4$ & $1.0$ & $0.61$ & -- & $1.7$ & $1.7$ & $1.3$ & $6.9$\\ 
Led. & $7$ & -- & $0.9$ & $0.54$ & -- & $1.7$ & $1.7$ & $1.7$ & $6.9$\\ 
Sch. & $7$ & -- & $0.9$ & $0.57$ & -- & $1.6$ & $1.6$ & $1.0$ & $6.9$\\ \hline
\end{tabular}}
\label{internal} 
\end{table}

\section{Convective criterion and internal structure of the models}
\label{internal_structure}

\begin{figure}[h!]
 \centering
 \includegraphics[scale=.45]{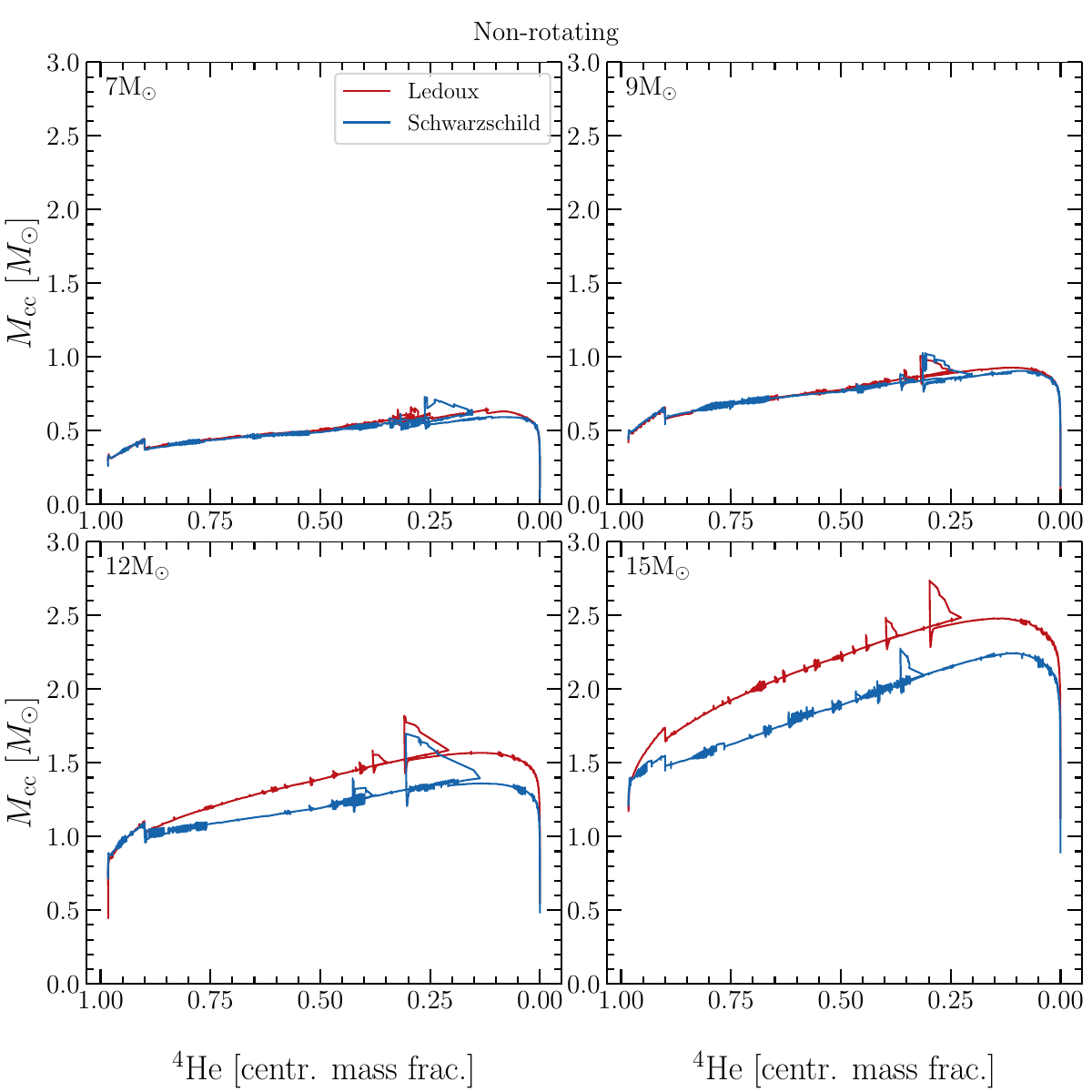}
 \caption{Relative mass of the convective core as a function of central He (from 1 to 0) during the He-burning phase for the non-rotating Ledoux (red) and Schwarzschild (blue) models between 7 and 15\,M$_\odot$.\\}
 \label{Mcc_Yc}
 \end{figure}

Changing the convective criterion impacts the sizes of the convective zones and thus the chemical structure of a star. Before discussing the impacts on the observable outputs of the stellar models using different criteria, we discuss here the direct impact on various internal properties of our stellar models. In Table~\ref{internal}, we present for each model the masses of the He- and CO cores at the end of the central H- and He-burning phases, respectively, the maximum extent (in Lagrangian mass coordinates) of the intermediate convective zone (ICZ) before the beginning of core He-burning, and the deepest reach of the outer convective zone (OCZ) at different evolutionary stages. 

We define the cores in the following way: the He core is the region inside the mass coordinate where at the first time the stellar layers are scanned from the surface towards the interior the mass fraction of helium $Y>0.9$. The boundary of the CO core is where for a similar scanning, it crosses $Y<10^{-2}$.\\

The masses of the He cores at the end of the MS phase obtained with the two different criteria show very small differences. For non-rotating models, the differences are always smaller than 3\%\footnote{We deduce this number by computing the Schwarzschild mass minus the Ledoux mass divided by the Ledoux mass.}. The same difference is obtained by comparing rotating models. These differences are blurred by other uncertainties pertaining to the models and are consistent with no effect due to the change in the convective criterion.
This is consistent with the fact that in phases during which the convective core size decreases in mass, no differences are expected when the Schwarzschild or the Ledoux criterion are applied because there is no mean molecular weight gradient just above the core. 

The only exception is the case of the rotating 120\,M$_\sun$ model. The core mass obtained with the  Schwarzschild criterion (33\,M$_\odot$) is much lower than the core mass obtained with the Ledoux model (55.6\,M$_\odot$). The structure evolution of the two rotating 120\,M$_\sun$ models shows intermediate convective shells that appear just above the core in the Schwarzschild models and quickly merge with the core. This grants more fuel to the core, prolongs the main sequence by $\sim$100\,kyr, and causes greater mass-loss of the star. 

Rotation increases the mass of the He core at the end of the MS phase. This increase is very similar for the Schwarzschild and Ledoux criteria. It is 5-6\% for the 9\,M$_\odot$, reaches a maximum for a mass about 40\,M$_\odot$ where it amounts to 26-27\%, and then decreases because mass loss by stellar winds becomes important and thus blurs the effects of rotation. The rise in the He-core mass with initial mass is linked to the higher efficiency of rotational mixing when the initial mass increases \citep[see the discussion in Sect.~3.1 of][]{MM2000AR}.

Figure~\ref{Mcc_Yc} shows how the mass of the convective core evolves during the core He-burning phase in the non-rotating stellar models for initial masses between 9 and 15\,M$_\odot$. The rotating cases (not shown here) exhibit very similar qualitative behaviours as those shown in Fig.~\ref{Mcc_Yc}: Their core masses increase during helium burning, and there are frequent breathing pulses\footnote{These breathing pulses are commonly agreed to be numerical artefacts and not of a physical nature.} that inject helium from the H-burning shell into the core. Interestingly, the convective cores are more massive in the 12 and 15\,M$_\odot$ models when the Ledoux criterion is applied. At first sight, this appears to be counter-intuitive. A stricter criterion for instability (Ledoux) should produce smaller convective cores than a less strict criterion (Schwarzschild). However, changing the criterion also affects the sizes of other convective zones in the models, and the changes in these other convective zones in turn affect the size of the convective core (see below). 

The CO-core masses (see Table~\ref{internal}) present modest differences for masses below and including 15\,M$_\odot$. The largest differences in the higher-mass star range are due to differences in the mass lost by stellar winds.

As expected, when the Schwarzschild criterion is used, larger intermediate convective zones associated with the H-burning shell appear. This has important consequences for the evolution of the He core, the occurrence of a blue loop \citep{MM2001}, and for the surface abundances of stars, which evolve back to the blue after having lost a large amount of mass during the red supergiant stage \citep{Saio2013, Georgy2014,Georgy2021}. The ICZ tends to increase the contribution of the H-burning shell to the total luminosity (as it transports energy more efficiently), and as a result, the star reacts to this by decreasing the energy generated in the He-burning core. The Schwarzschild core has a lower central temperature than the Ledoux core and is therefore less dense. The Schwarzschild core has the same extent in radius as the Ledoux core, but its mass is lower, as shown in the bottom panels of Fig~\ref{Mcc_Yc}.

A convective envelope does not appear in the high-mass range (above 60\,M$_\odot$) as a result of mass loss, which prevents the stars from evolving in the red part of the Hertzsprung-Russell diagram (HRD). The outer convective zones for the 40\,M$_\odot$ models are very small. The 32\,M$_\odot$ initial mass is a transition case, without a convective envelope for the rotating models and with a significant envelope, reaching down to 25.5\,M$_\odot$ for Schwarzschild and 16\,M$_\odot$ for Ledoux, in the non-rotating case. 
For masses below and including 25\,M$_\odot$, all the models have significant convective envelopes during the core He-burning phase. Models between 12 and 25\,M$_\odot$ computed with the Ledoux criterion have deeper outer convective zones, but this is not the case for the 7 and 9\,M$_\odot$ (see Sect.~\ref{blueloops}). 

It might appear surprising that adopting a criterion that hinders convection can increase the extent of the outer convective zone. This extent also strongly depends on the position of the star in the HRD during the core He-burning phase, however: A hotter position during that phase decreases the extent of the outer convective zone. The position in the HRD at which core He-burning occurs is very sensitive to the variation in the abundances above the core as well as to the presence or absence of an intermediate convective zone. The intermediate convective zone typically tends to make the star more compact (i.e. blue), which then reduces the size of the convective envelope. In this manner, a criterion favouring convection (Schwarzschild) can produce a larger ICZ and consequently a less extended OCZ.\\

\section{Evolutionary tracks and lifetimes}
\label{Evolution}
% HRD WITH CENTRAL HELIUM MASS FRACTION
 \begin{figure*}[h!]
 \centering
 \includegraphics[scale=1]{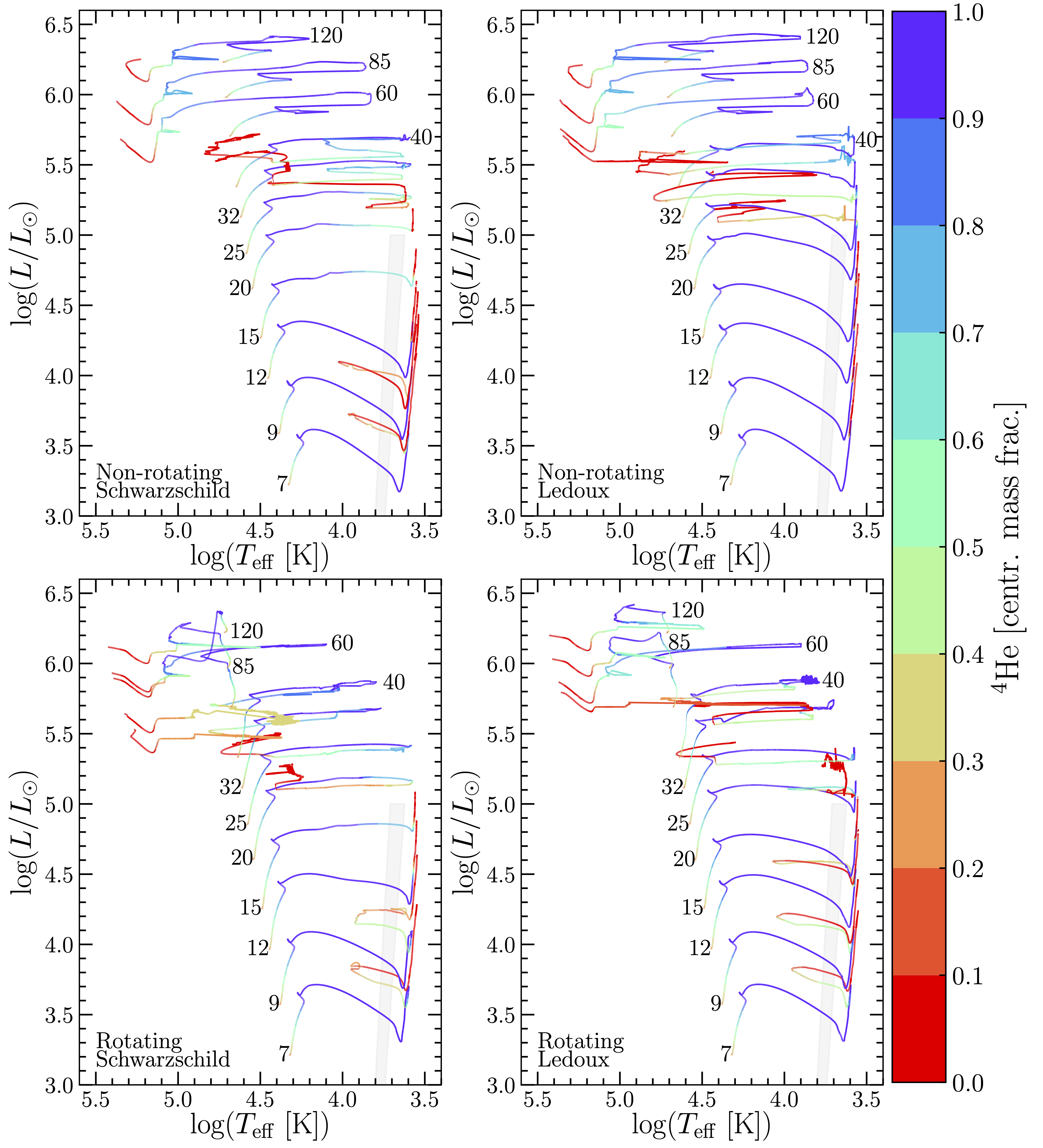}
 \caption{Evolutionary tracks of our stellar models in the HRD, colour-coded by the central helium mass fraction $Y_c$. The grey shaded area represents the Cepheid instability strip \citep{Tammann2003}.}
 \label{HRandYC}
 \end{figure*}

We show the evolutionary tracks for all models in Fig.~\ref{HRandYC}. 
% MS
Main-sequence tracks are as expected to be similar in the Schwarzschild and Ledoux models. The main difference occurs in the rotating 120\,M$_\odot$ models, as mentioned in Sect.~\ref{internal_structure}.

%Post-MS difference below and above 40M
After the MS phase, the largest differences between the tracks (whether rotating or not) computed with the Schwarzschild and Ledoux criteria occur for the mass range between 7 and 40\,M$_\sun$. In the upper mass range (between 60 and 120\,M$_\sun$), stellar winds are the main factor governing the evolution of the stars. In the next paragraphs, we discuss in more detail how changing the convection criterion affects the first crossing of the Hertzsprung gap (Sect.~\ref{first-crossing}), the blue loops (Sect.~\ref{blueloops}), the properties of the Cepheids (Sect.~\ref{cepheids}), the properties of stars ending their evolution as red supergiants (Sect.~\ref{RSG}), and the properties of stars ending their evolution as yellow (YSG), blue supergiants (BSG), or as Wolf-Rayet stars (Sect.~\ref{YSG-BSG-LBV-WR}).

\subsection{First crossing of the Hertzsprung gap}
\label{first-crossing}
% Teff vs Yc, central He burning 7-15Msol
\begin{figure*}[h!]
 \centering
 \includegraphics{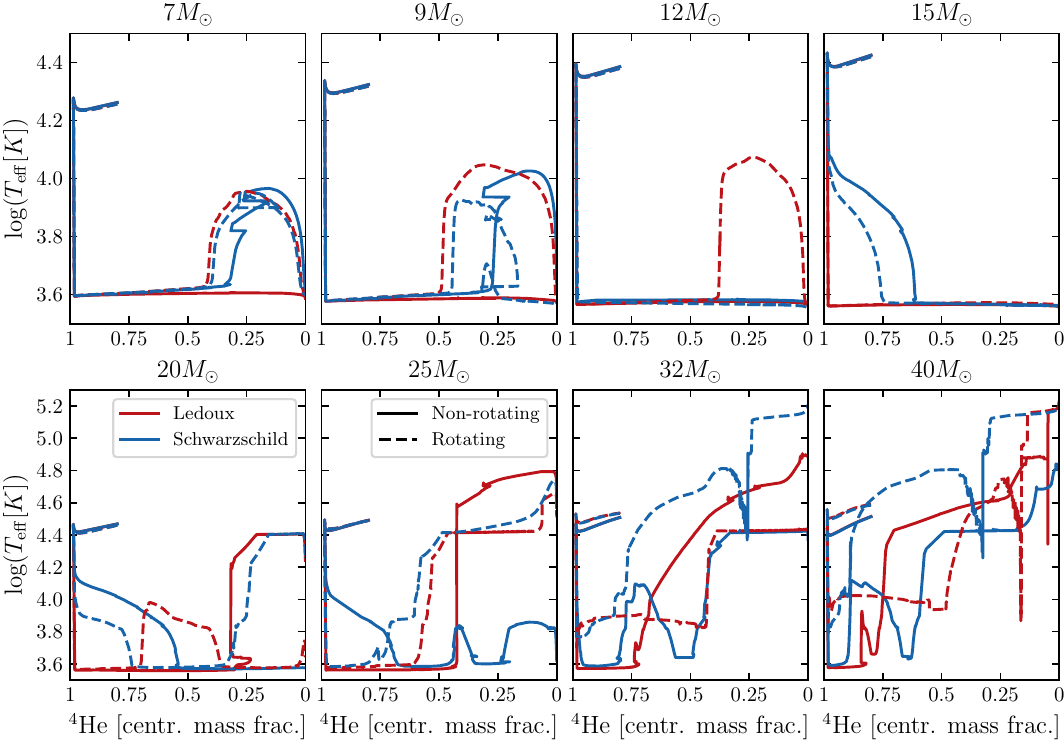}
 \caption{Effective temperature towards the end of the main sequence ($Y_{\rm c}>0.8$) and core He-burning for non-rotating (solid curves) and rotating (dashed curves) Ledoux (red) and Schwarzschild (blue) models. Upper row: Masses between 7 and 15\,M$_\odot$. Lower row: Masses between 20 and 40\,M$_\odot$. In each row, the vertical axes in all panels share the same limits, which are indicated in the leftmost panels.}
 \label{YTeff}
\end{figure*}
\begin{figure*}[h!]
    \centering
    \includegraphics{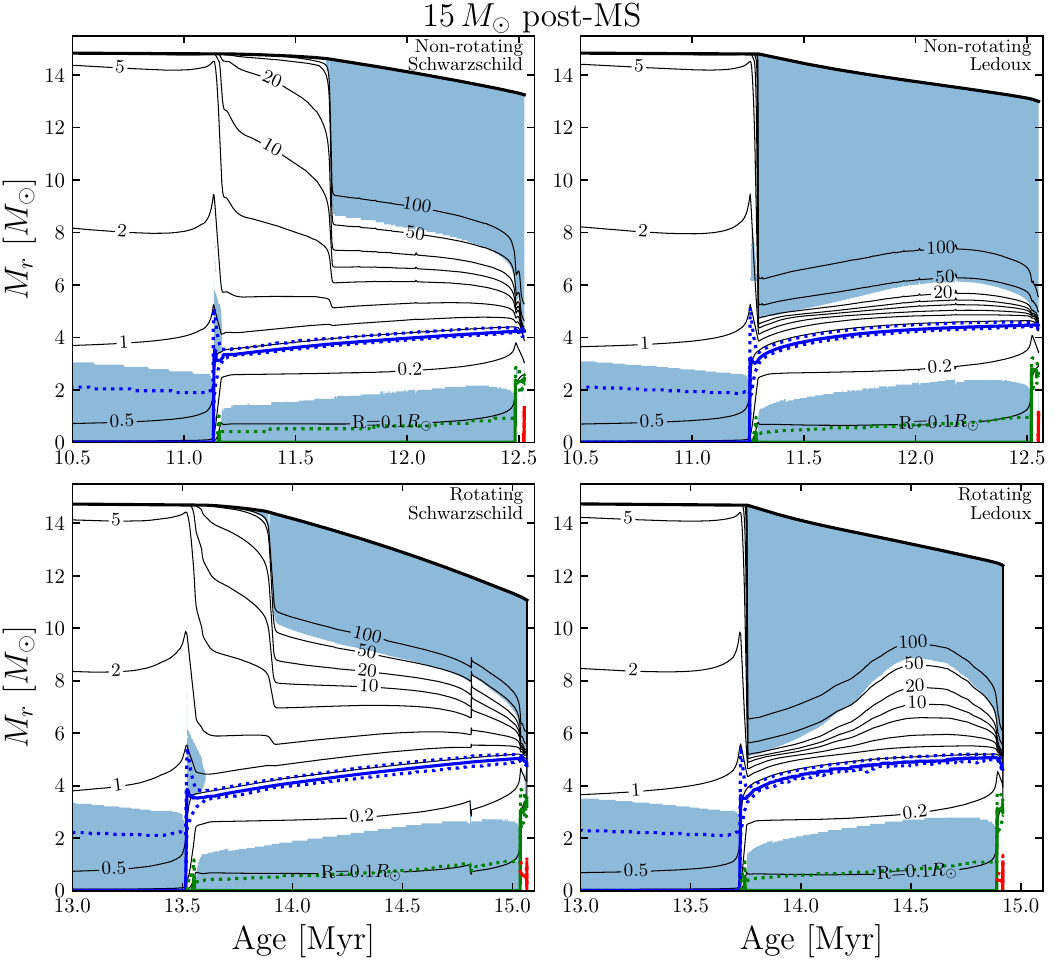}
    \caption{Kippenhahn diagrams showing the post-MS phases for the four models computed with an initial mass of 15\,M$_\odot$. The vertical axis shows the mass coordinate, and the numbered lines show isoradius levels. Left column: Schwarzschild models. Right column: Ledoux models. Upper row: Non-rotating models. Lower row: Rotating models. The intermediate convective zone above the H-burning shell is clear in the Schwarzschild models and is associated with the slower expansion of the envelope.}
    \label{Kipp_15_PMS}
\end{figure*}

Figure~\ref{YTeff} shows the evolution of the effective temperature as a function of the mass fraction of helium at the centre $Y_c$ for all the models with initial masses between 7 and 40\,M$_\odot$.

Models with masses below and including 12\,M$_\odot$ all begin their core He-burning phase when the star is a red supergiant, with $\log {(T_{\rm eff}~\rm{[K]})} \sim 3.6$.
For the 15-25\,M$_\odot$ mass range, the Ledoux criterion favours a quick crossing of the Hertzsprung gap, whether the models are computed with or without rotation. This is likely to be related to the fact that the use of the Ledoux criterion restrains the apparition and/or the extension of the intermediate convective zone associated with the H-burning shell, as shown in Fig.~\ref{Kipp_15_PMS}, which shows Kippenhahn diagrams for the post-MS phases of the four 15\,M$_\odot$ models. This tends to produce steeper gradients of H and He in that zone and favours helium ignition in the core in the red supergiant stage  \citep{Farrell2022}.
The models of 15 and 20\,M$_\odot$ with the Schwarzschild criterion (as well as the non-rotating 25\,M$_\odot$ model) show a different behaviour. These models ignite helium in their cores when the star still has a high effective temperature ($\log{T_{\rm eff}~\rm{[K]}} > 4.0$), whether it rotates or not. This is linked to the larger intermediate convective zone in these models when the Schwarzschild criterion is used. A tentative explanation is that convection is more effective at transporting energy than radiation, and therefore, the presence of this ICZ allows the increased luminosity generated by the contraction of the core to reach the surface more easily. In the Ledoux models, however, energy is transported by radiation, and a larger fraction of it is deposited into the envelope itself, causing its rapid expansion and a drop in its luminosity. In both cases, the core contraction timescales are similar; the ICZ only affects the timescale of the envelope expansion. This means that at the onset of core helium burning, the Schwarzschild star remains in a blue part of the HRD.\\
The cases of the more massive models (equal to or higher than 25\,M$_\odot$) result from intricate interactions between convection, rotation, and mass loss. In general, helium ignition in the core of these models always occurs at an effective temperature below
$\log{T_{\rm eff}~\rm{[K]}}=4.0$. The only exception is the non-rotating 25\,M$_\odot$ Schwarzschild model. 

The evolution of the luminosity during the first crossing varies depending on the convective criterion.
In general, the Ledoux models for masses between 15 and 20\,M$_\odot$ experience a stronger decrease in luminosity before going up the Hayashi line than Schwarzschild models because these models are in a stronger radiative disequilibrium than the Schwarzschild models. Because they expand more rapidly, a larger amount of energy per unit time is tapped from the gravitational contraction of the core to expand the envelope. This tends to decrease the amount of energy that is radiated away. This decrease is to some extent a consequence of the time taken to cross the Hertzsprung gap. If it occurs on a sufficiently long timescale, the decrease in luminosity is modest. When it occurs on a rapid timescale, the luminosity drop is marked. To summarise, the Schwarzschild criterion favours the apparition of an ICZ above the H-burning shell after the main sequence, and for the reasons already described just above, this favours He ignition in the core when the star is still in the blue part of the HRD. The crossing of the Hertzsprung gap occurs on a longer timescale (nuclear, about 500\,kyr) than with the Ledoux criterion (thermal, about 30\,kyr). With the Schwarzschild criterion, the envelope has more time to radiate
the excess energy produced by the contraction of the core away, and the Hertzsprung gap crossing
occurs at roughly constant luminosity. The Ledoux criterion suppresses the ICZ, implying a shorter timescale for the Hertzsprung gap crossing. A large part of the energy radiated by the core contraction is absorbed by the envelope. As a result, the luminosity at the surface decreases significantly during the crossing.

\subsection{Blue loops}
\label{blueloops}
%He4 profiles around the blue loop, non-rotating 7Msol
 \begin{figure}
 \centering
 \includegraphics{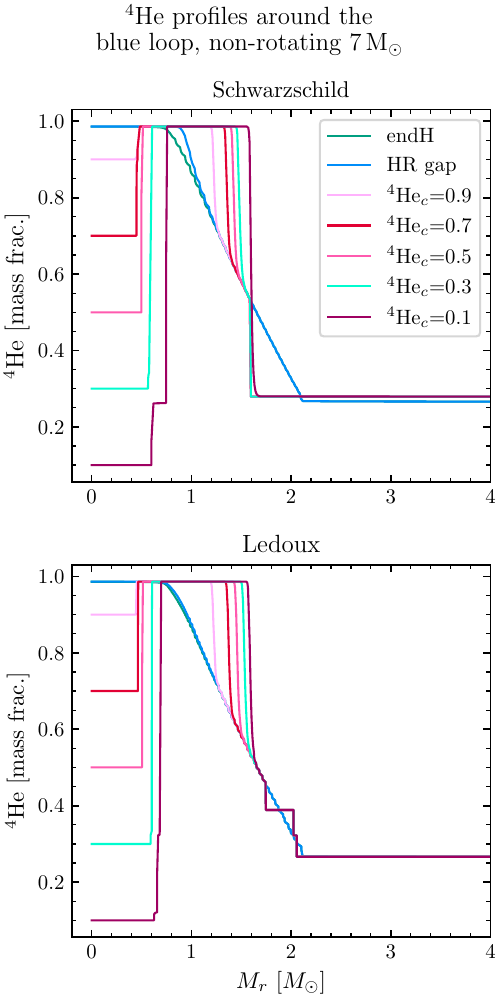}
 \caption{$^4$He profiles at different times before and during core He-burning for the two non-rotating 7\,M$_\odot$ models.}
 \label{P007_Heprofiles_BL}
 \end{figure}

Fig.~\ref{YTeff} shows that for initial masses above 20\,M$_\odot$ the effective temperature can increase during core He-burning. These phases are not considered as blue loops for two reasons, however. First, these stars end their lifetimes at high effective temperatures as BSG or WR stars. Second, this evolution is mainly driven by mass loss and not by some internal evolution due to changes in the hydrostatic structure. In the rest of this section, we thus focus on the mass domain between 7 and 15\,M$_\odot$.\\

For the lower-mass models, blue loops, characterised by an increase in effective temperature (blueward evolution) around $Y_c\sim0.4-0.3$, appear and are followed by a decrease (redward evolution) at the end of core He-burning. The physics of the blue loops was discussed in detail in previous papers \citep[see e.g.][]{Laut1971,MM2001, Anderson2014}. Many aspects of the stellar models influence
the blue loops: for instance the rate of the $^{14}N(p,\gamma)^{15}O$ reaction \citep{N142012},
of the $^{12}C(\alpha,\gamma)^{16}O$ reaction \citep{Brunish1990},
the undershooting below the outer convective zone \citep{Lai2011}, the helium and metal content \citep{Bono2000}, and rotation \citep{Zhao2023}. Recently, the occurrence of blue loops was used to constrain the magnetic moment of massive neutrinos \citep{neutri2020} or the impact of axions \citep{Axions2013}. Stellar models accounting for these effects undergo an additional loss of energy, and this tends to suppress the blue loops. This demonstrates that if the evacuation of energy from the core is facilitated, then this favours the suppression of the blue loops. Here, we focus on only one aspect, namely the impact of the convective criterion.

It has been shown that the blue loops may be linked to subtle differences in the abundance distributions near the H-burning shell, and that excess helium above the shell will suppress the blue loop \citep[see e.g.][who propose that the increase in mean molecular weight caused by having helium instead of hydrogen above the shell is the main culprit in suppressing the blue loop]{Walmswell2015}. \citet{Laut1971} have proposed a criterion based on the gravitational potential of the core in order to decide whether a model produces a blue loop. Stars whose core has a gravitational potential higher than a given critical value remain along the Hayashi line, while those whose core has a gravitational potential lower than this critical value develop a blue loop.
Interestingly, the quantity to be compared to the critical value is the core potential only if there is a steep chemical gradient just above the He core or near the H-burning-shell. Otherwise, the gravitational potential of the core has to be multiplied by a factor that is larger for a milder gradient, thus inducing the model to stay along the Hayashi line.

%blue loops 7-9
The non-rotating 7 and 9\,M$_\odot$ models computed with the Ledoux criterion show no blue loops, while the 7 and 9\,M$_\odot$ models with the Schwarzschild criterion have well-developed loops. Because these two masses behave very similarly in the features of their overall evolution, we focus on the 7\,M$_\odot$ models, and the results qualitatively apply to the 9\,M$_\odot$ models. Figure~\ref{P007_Heprofiles_BL} shows the $^4$He profiles at different times before and during core He-burning for the two non-rotating 7\,M$_\odot$ models. In the Schwarzschild model, the outer convective zone has equalised the abundances from the surface down to a mass near 1.6\,M$_\odot$, reducing the extent of the zone with a chemical composition gradient (see the line corresponding to $^4$He$_{c}$ where the gradient is between the mass coordinates  1.3 and 1.6\,M$_\odot$). This occurs at the very beginning of the core He-burning phase. In the Ledoux model, the base of the outer convective zone always remains above the mass coordinate of 2.1\,M$_\odot$ and the zone of the chemical composition gradient is larger (from 1.3 to 2.1\,M$_\odot$), thus milder, than in the Schwarzschild model. This acts to prevent the formation of a blue loop. The Ledoux model also has an excess of helium above the hydrogen-burning shell (at 1.7-2.1\,M$_\odot$), which suppresses the blue loop \citep[see again][]{Walmswell2015}. The Ledoux models have this excess helium but the Schwarzschild models lack it because the mean molecular weight gradient above the hydrogen-burning shell stabilises this region against convection when the Ledoux criterion is used. As a result, the convective envelope does not reach down far enough to equalise the helium abundance to the surface value, leaving a helium excess above the shell.

Models of 7 and 9\,M$_\odot$ with rotation present less marked differences when the convective criterion is changed from the Schwarzschild to the Ledoux criterion. Rotational mixing blurs the differences that arise from different choices of the criterion for convection, and chemical composition gradients as well as helium abundances above the H-burning shell are similar for those models.

For the non-rotating 12\,M$_\odot$, we do not observe any blue loop or any difference between the Schwarzschild and the Ledoux models.
The only 12\,M$_\odot$ model showing a blue loop is the rotating Ledoux model. This model would thus predict the most luminous Cepheids of those discussed in this paper (excluding the 15\,M$_\odot$ models that cross the Cepheid instability strip very briefly).\\

\subsection{Cepheid properties}
\label{cepheids}

% CEPHEID DURATIONS
 \begin{figure*}[h!]
 \centering
 \includegraphics[scale=.5]{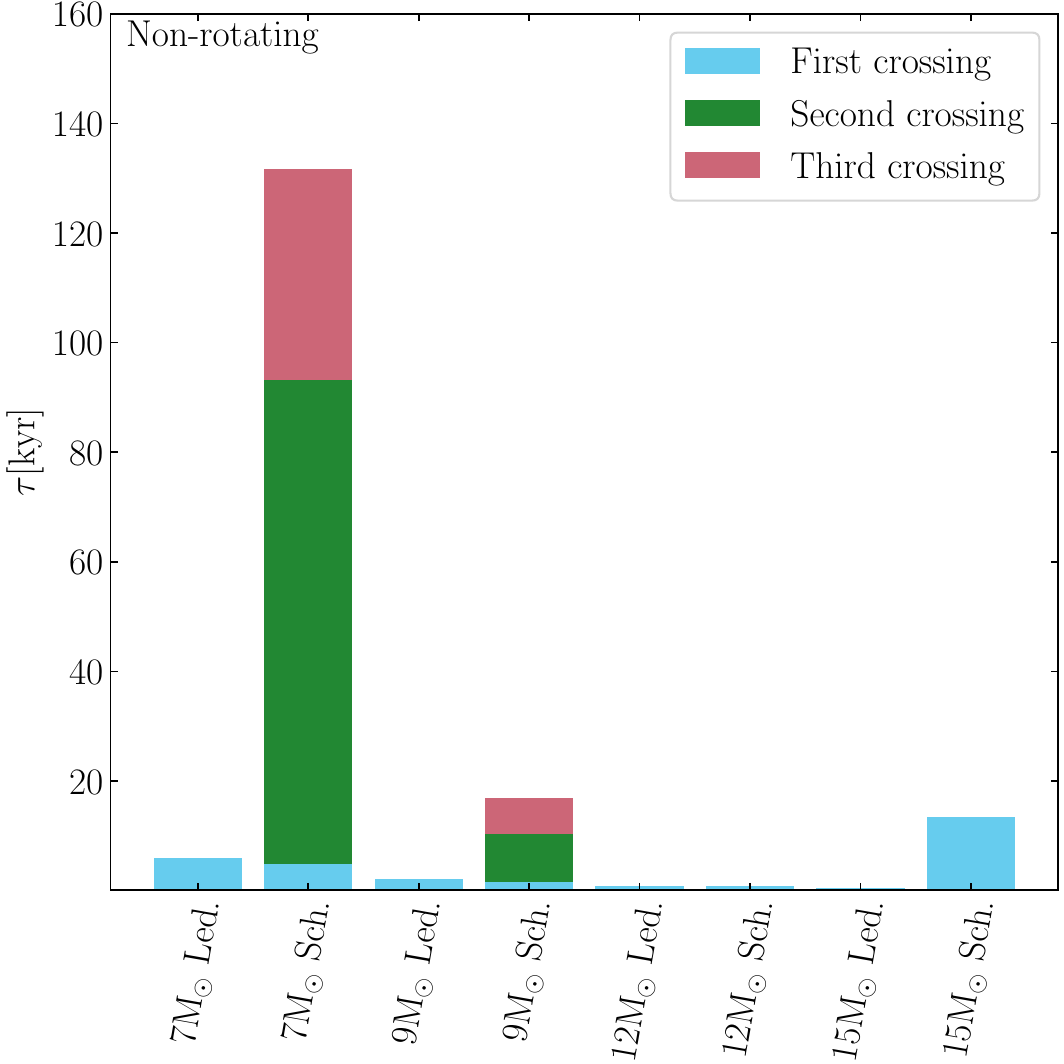}\includegraphics[scale=.5]{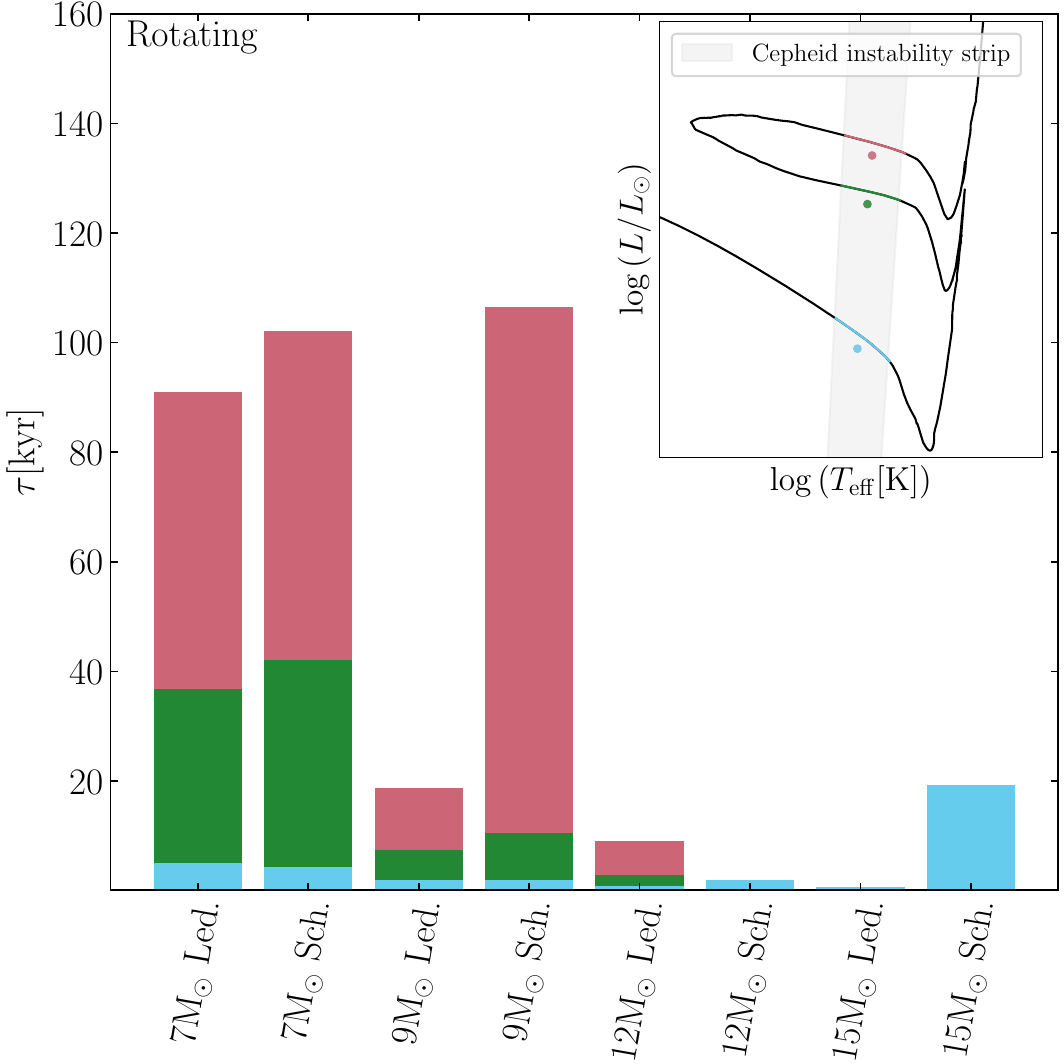}
 \caption{Bar plot showing the durations of the different Cepheid phases for masses at which at least one of the four (Ledoux, Schwarzschild, rotating, non-rotating) models has a Cepheid phase. We distinguish between the first crossing (redward crossing of the Hertzsprung gap after the MS), the second crossing (blueward during the blue loop), and the third crossing (redward during the blue loop). These three moments are indicated in the small top right inset by coloured lines and dots with corresponding colours. Left panel: Non-rotating model. Right panel: Rotating model.}
 \label{Ceph_durations}
 \end{figure*}

% CEPHEID N ABUNDANCES
 \begin{figure*}[h!]
 \centering
 \includegraphics[scale=.5]{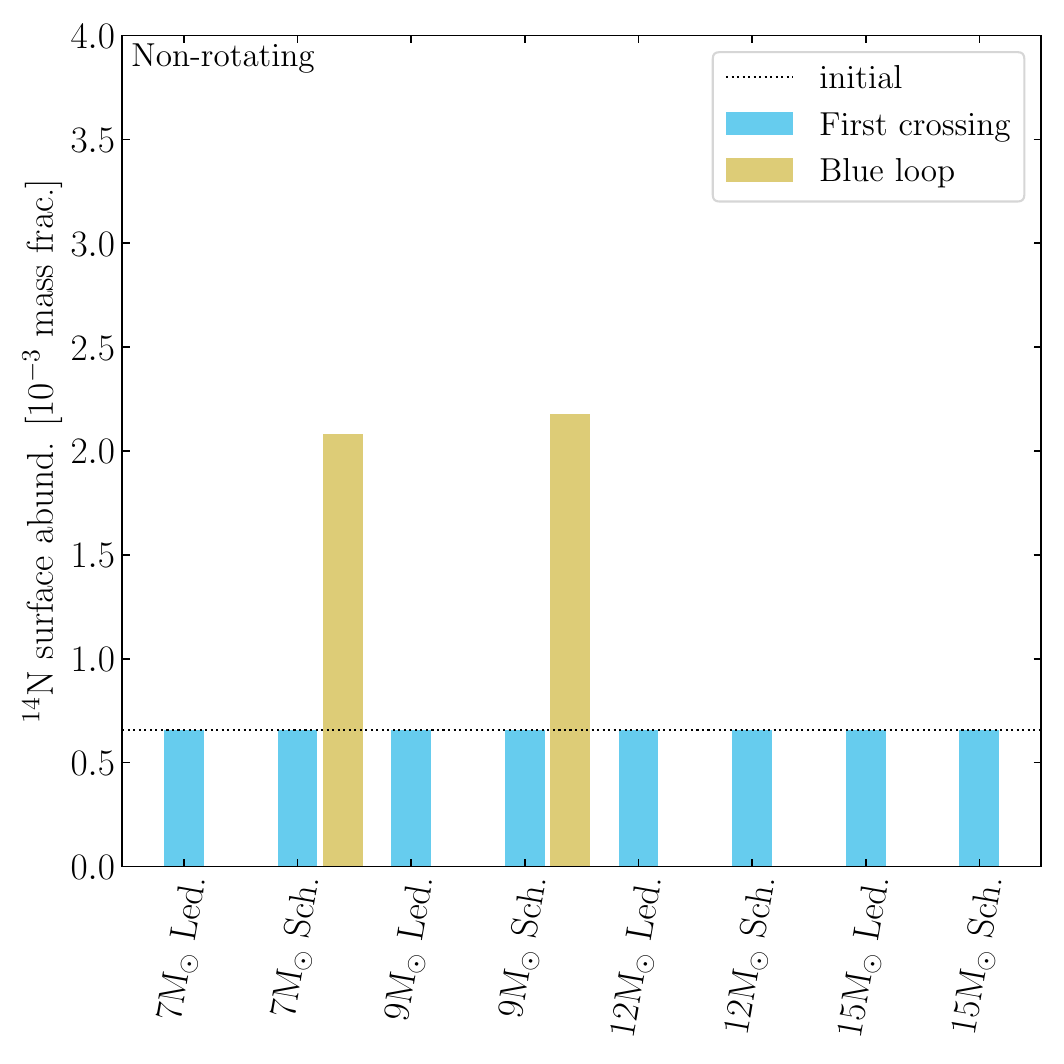}\includegraphics[scale=.5]{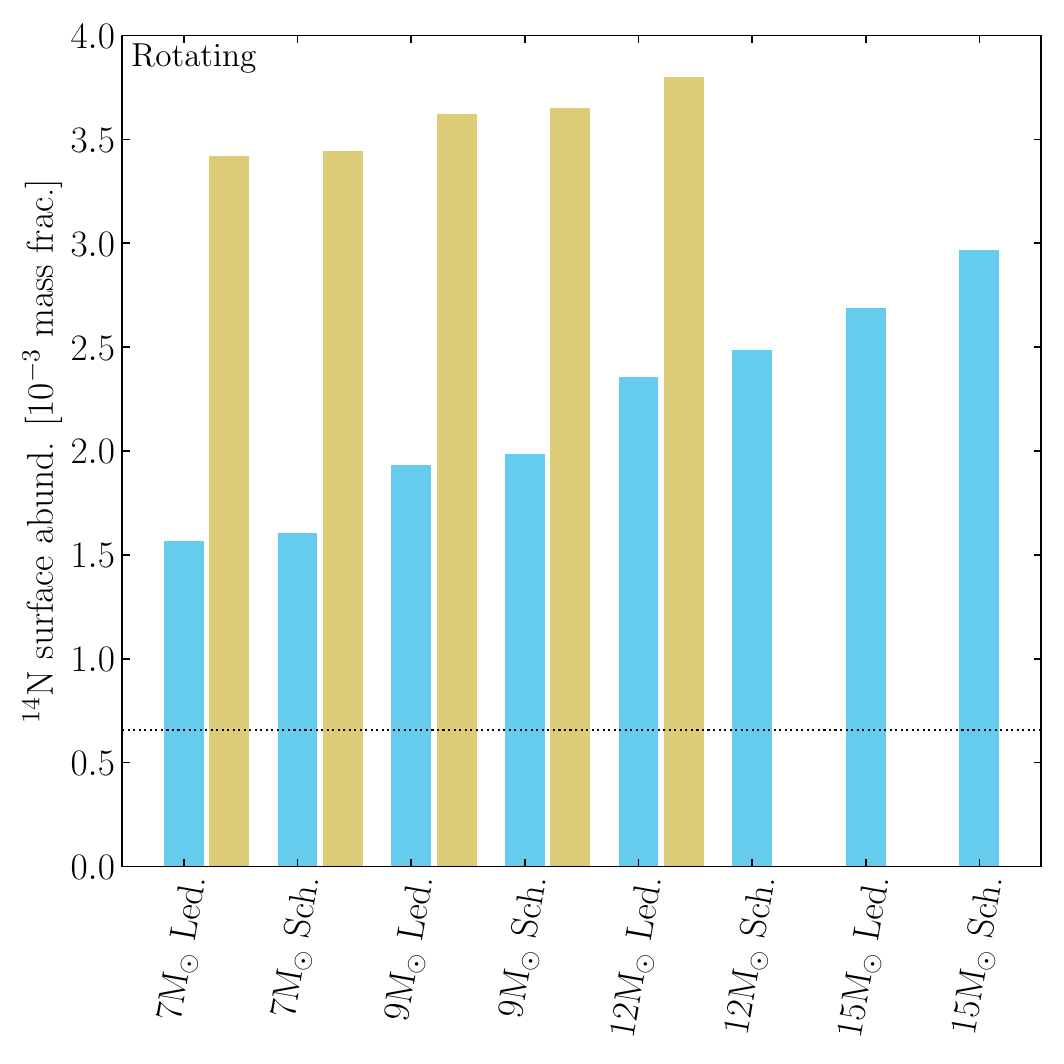}
 \caption{Bar plot showing the surface abundances of $^{14}\rm{N}$ before (cyan bar) and during (yellow bar) the blue loop for stars between 7 and 15\,M$_\odot$. The dotted line indicates the initial surface abundance. Left panel: Non-rotating model. Right panel: Rotating model.}
 \label{Ceph_Nsurf}
 \end{figure*}

% CEPHEID SURFACE VELOCITIES
 \begin{figure*}
 \centering
 \includegraphics[scale=.5]{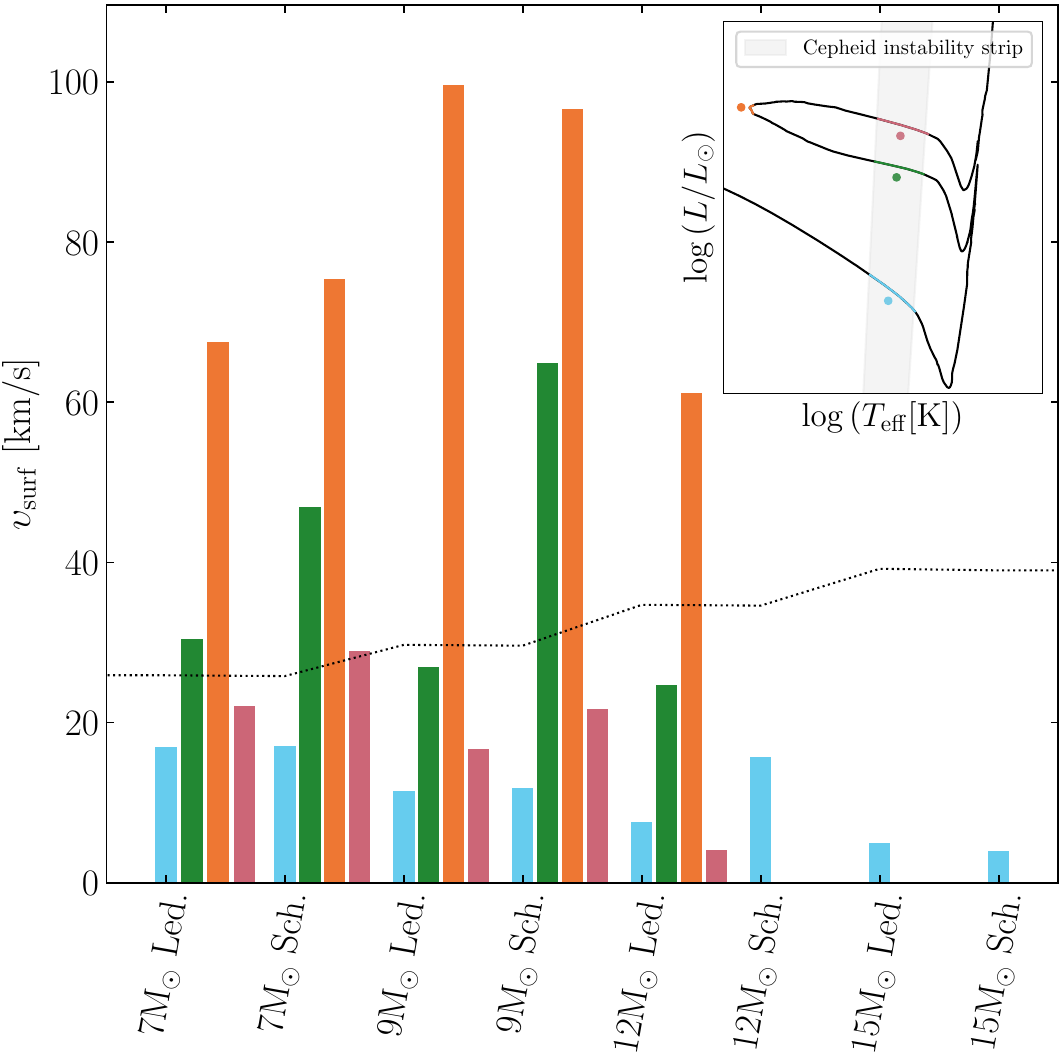}\includegraphics[scale=.5]{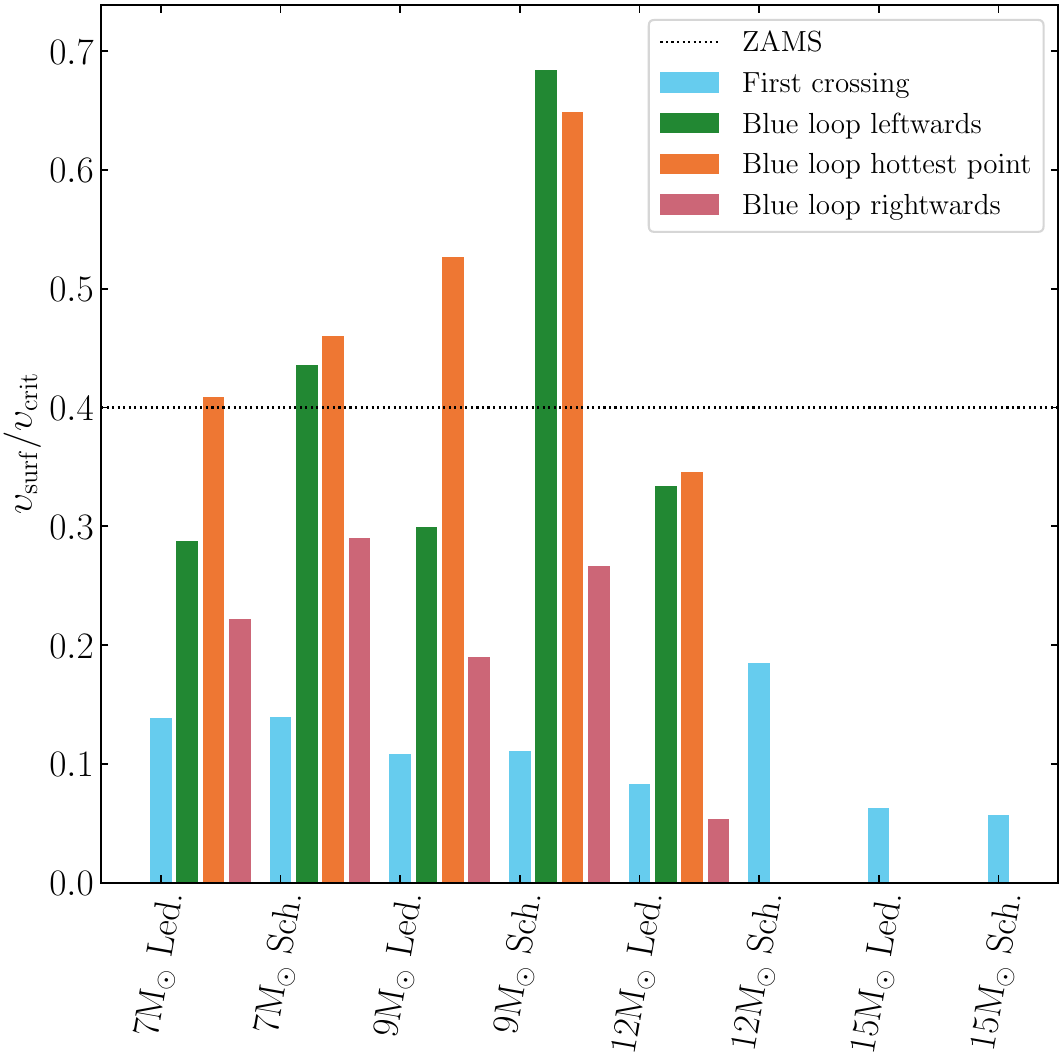}
 \caption{Bar plot showing the surface velocities of rotating stars during the first crossing of the Hertzsprung gap and at three different times of the blue loop for stars between 7 and 15\,M$_\odot$. These four moments are indicated in the small top right inset by coloured lines and dots with corresponding colours. Left panel: Equatorial velocity $\upsilon_{\rm surf}$. Right panel: Ratio of the equatorial to the critical velocity $\upsilon_{\rm surf}/\upsilon_{\rm crit}$.}
 \label{Ceph_Vsurf}
 \end{figure*}

Here, we discuss results concerning stars that cross the Cepheid instability strip (shown as the grey shaded region in Fig.~\ref{HRandYC} and in the small panels of Figs.~\ref{Ceph_durations} and \ref{Ceph_Vsurf}). The initial masses of these stars are between 7 and 15\,M$_\odot$.

Figure~\ref{Ceph_durations} shows the durations of the different Cepheid phases. As is already well known, the first crossing is very short in all cases: It is much shorter than the blue loop when one occurs. As a result, most observed Cepheids should be undergoing a blue loop rather than crossing the Hertzsprung gap for the first time. For non-rotating models, the Schwarzschild criterion in general favours a longer duration of the Cepheid phase. The 7 and 9\,M$_\odot$ Schwarzschild models undergo blue loops and the Ledoux models do not. Moreover, the 15\,M$_\odot$ Schwarzschild models start core helium-burning at an effective temperature of $\log{(T_{\rm eff} [K])}\sim4$ and thus subsequently evolve on a helium nuclear timescale, whereas the Ledoux stars begin core helium-burning as red supergiants and thus cross on a much shorter Kelvin-Helmholtz timescale.\\
For rotating models, the situation is less different between the Schwarzschild and Ledoux models, although the case of the 9\,M$_\odot$ shows very significant differences: the Schwarzschild model spends 110\,kyr as a Cepheid, while the Ledoux model spends only 20\,kyr. We find no obvious correlation between the overall duration of the Cepheid phase and the stellar mass.

Figure~\ref{Ceph_Nsurf} shows the surface abundances of $^{14}\rm{N}$ before and during the blue loop. In non-rotating models, there is no nitrogen surface enrichment during the MS phase, and as a result, no model shows nitrogen enrichment during the first crossing. Non-rotating models show an increase in surface nitrogen when they go through the RSG stage, where some nitrogen is dredged up to the surface through convection. This dredge-up occurs before the blue loop and is thus visible during the blue loops for the 7 and 9\,M$_\odot$ models. For rotating models, the impact of rotational mixing during the MS is apparent in the surface $^{14}\rm{N}$ during the first crossing as it is larger than the initial value for all models.
Interestingly, in the nitrogen surface abundance of the 9\,M$_\odot$ Schwarzschild models, the blue-loop value obtained for the non-rotating model is equivalent to that given by the rotating model before the blue loop. This shows that the surface nitrogen abundance by itself is a poor indicator of the stellar phase (before or after the blue loop), unless its surface velocity is measured (see more on this point below).
The nitrogen surface abundance increases with the initial mass, and it is slightly higher for Schwarzschild than for Ledoux models. It is higher in rotating models than in non-rotating models during the blue loop. 

Figure~\ref{Ceph_Vsurf} shows the surface velocities of rotating stars at the three moments described above, as well as at the hottest point of the blue loop. In the 9\,M$_\odot$ Schwarzschild model, the surface rotation speed along the blue loop strongly increases during blueward evolution (from 10~km/s up to nearly 100~km/s or nearly 70\% of the critical velocity). Interestingly, it decreases very fast during the redward evolution after reaching the hottest point of the blue loop. The strong surface acceleration arises because during the RSG phase, an extended convective envelope is present. This convective zone rotates as a solid body, which means that the angular momentum is highest at the outer border of the convective zone. When the star contracts, the angular momentum accumulated by convection in the outer layers produces a rapid acceleration. The star loses angular momentum by winds during the blue loop, and it has a larger radius after than before (the effective temperature is similar, but the luminosity has increased). As a result, when it evolves back to the red, its situation is not the same as before the blue loop: Its surface velocity is lower. The Ledoux model shows a qualitatively similar behaviour, but the different evolution of the outer convective zone makes the evolution of the surface rotation slightly different.

\subsection{Red supergiants}
\label{RSG}

In this and the next section, we examine the impact of the convective criterion on different subclasses of post-MS stars. These subclasses are defined in Table~\ref{TabSubtypes}.

%DEFINITION OF STELLAR SUBTYPES
\begin{table}[h!]
\caption{Definitions of stellar subtypes in populations.}
\centering
\scalebox{0.9}{
\begin{threeparttable}
\begin{tabular}{|c|c|}
\hline \hline
Subtype & Definition\\
\hline
MS & $X_c\tnote{$\dagger$} ~\geq 10^{-5}$ \\
Post-MS & $X_c < 10^{-5}$ \\
SG & Post-MS \& $\log{(L~[L_\odot])}>4$ \& $X_s\tnote{$\dagger$}~ > 0.35$\\
BSG & SG \& $\log{(T_{\rm eff}~[\rm{K}])}\geq3.9$\\
YSG & SG \& $3.66<\log{(T_{\rm eff}~[\rm{K}])}< 3.9$ \\
RSG & SG \& $\log{(T_{\rm eff}~[\rm{K}])}\leq3.66$\\
WR & $\log{(T_{\rm eff}~[\rm{K}])}>4$ \& $X_s \leq 0.35$ \\
WNL & WR \& $X_s \geq 10^{-5}$ \\
WNE & WR \& $X_s < 10^{-5}$ \& $C_s\tnote{$\dagger$}~ \leq $$N_s$\tnote{$\dagger$} \\
WNC & WR \& $X_s < 10^{-5}$ \& $0.1 \leq$ $C_s/N_s \leq 10 $ \\
WC & WR \& $X_s < 10^{-5}$ \& $C_s$ + $O_s\tnote{$\dagger$}~\leq$ $Y_{\rm s}\tnote{$\dagger$}~$\,(numb. frac.)\\
WO & WR \& $X_s < 10^{-5}$ \& $C_s$ + $O_s$ > $Y_{\rm s}$ (numb. frac.) \\
\hline 
\end{tabular}
\begin{tablenotes}
\item [$\dagger$] $X$, $Y$, $C$, $N$, and $O$ refer to the mass fractions (unless `numb. frac.' is specified, in which case they are the number fractions) of $^1$H, $^4$He, $^{12}$C, $^{14}$N, and $^{16}$O, respectively. The subscripts $c$ (centre) and $s$ (surface) refer to where in a star these abundances are taken.
\end{tablenotes}
\end{threeparttable}}
\label{TabSubtypes} 
\end{table} 

Here, we discuss whether the convective criterion affects the maximum luminosity of red supergiants.
Fig.~\ref{HRandYC} shows that the maximum luminosity reached for red supergiants is not much affected by the convective criterion, but it is affected by rotation. Rotating models produce an upper limit for the RSG luminosity of about $\log{(L~[L_\odot])}=5.4$, while the non-rotating models extend this upper limit to values of about $\log{(L~[L_\odot])}=5.7$.

When we now compare the upper luminosity of red supergiants that remain red until the end of their evolution and thus will be progenitors of a type II SN, then the upper limits are lower than those in the previous paragraph because mass loss by stellar winds induces a blueward evolution and thus produces yellow or blue progenitors in the higher-luminosity range (see Sect.~\ref{YSG-BSG-LBV-WR}). This upper limit in luminosity for red supergiant core-collapse progenitors is about $\log{(L~[L_\odot])}=5.2$ for the non-rotating Schwarzschild models, $\log{(L~[L_\odot])}=4.95$ for the non-rotating Ledoux models, $\log{(L~[L_\odot])}=5.1$ for rotating Schwarzschild models, and $\log{(L~[L_\odot])}=5.05$ for the rotating Ledoux models. 
We note that the models that spend a larger fraction of their core He-burning phase in the red supergiant phase have slightly lower upper limits. This is expected because their mass is reduced by strong RSG winds for a longer duration. This tends to lower the minimum mass above which stars later evolve blueward. For instance, in the non-rotating case (with the largest difference in maximum RSG luminosity), the most luminous Schwarzschild RSG has an initial mass of 20\,M$_\odot$ while for the Ledoux RSG, the initial mass is 15\,M$_\odot$ because the 20\,M$_\odot$ model becomes blue after its mass-loss episode.

\subsection{Yellow and blue supergiants, luminous blue variables, and Wolf-Rayet stars}
\label{YSG-BSG-LBV-WR}

% BIG TABLE
\begin{table*}[h!]
\caption{Properties of the stellar models at the end of the H-, He-, and C- burning phases.}
\centering
\scalebox{.65}{\begin{tabular}{|ccccc|cccccc|cccccc|cccccc|}
\hline\hline
\multicolumn{5}{|c|}{} & \multicolumn{6}{c|}{End of H-burning} & \multicolumn{6}{c|}{End of He-burning} & \multicolumn{6}{c|}{End of C-burning}\\ 
Crit. & M$_\text{ini}$ & $\upsilon_\text{ini}$ & $\upsilon_\text{ini}/\upsilon_\text{crit}$ & $\bar{v}_\text{MS}$ & $\tau_\text{H}$ & M & $\upsilon_\text{surf}$ & $Y_\text{surf}$ & $\text{N}/\text{C}$ & $\text{N}/\text{O}$ & $\tau_\text{He}$ & M & $\upsilon_\text{surf}$ & $Y_\text{surf}$ & $\text{N}/\text{C}$ & $\text{N}/\text{O}$ & $\tau_\text{C}$ & M & $\upsilon_\text{surf}$ & $Y_\text{surf}$ & $\text{N}/\text{C}$ & $\text{N}/\text{O}$ \\ 
& $M_{\sun}$ && \multicolumn{2}{c|}{km s$^{-1}$} & Myr & $M_{\sun}$ & km s$^{-1}$ & \multicolumn{3}{c|}{mass fract.} & Myr & $M_{\sun}$ & km s$^{-1}$ & \multicolumn{3}{c|}{mass fract.} & kyr & $M_{\sun}$ & km s$^{-1}$ & \multicolumn{3}{c|}{mass fract.}\\ 
\hline
Led. & $120$ & $0.4$ & $397$ & $120$ & $3.140$ & $64$ & $3.5$ & $0.93$ & $78.11$ & $86.75$ & $0.305$ & $30.9$ & $10.1$ & $0.24$ & 0 & 0 & $0.010$ & $30.8$ & $21.7$ & $0.24$ & 0 & 0 \\ 
Sch. & & $0.4$ & $389$ & $110$ & $3.178$ & $34.7$ & $0.6$ & $0.97$ & $70.21$ & $92.04$ & $0.350$ & $19.1$ & $2$ & $0.27$ & 0 & 0 & $0.030$ & $19$ & 0 & $0.26$ & 0 & 0 \\ 
Led. & & -- & -- & -- & $2.664$ & $63.3$ & -- & $0.78$ & $95.88$ & $77.06$ & $0.311$ & $28.7$ & -- & $0.23$ & 0 & 0 & $0.012$ & $28.6$ & -- & $0.23$ & 0 & 0 \\ 
Sch. & & -- & -- & -- & $2.672$ & $63.7$ & -- & $0.78$ & $94.93$ & $77.28$ & $0.307$ & $31$ & -- & $0.24$ & 0 & 0 & $0.008$ & $30.9$ & -- & $0.24$ & 0 & 0 \\ \hline
Led. & $85$ & $0.4$ & $372$ & $133$ & $3.698$ & $43.4$ & $6.1$ & $0.97$ & $68.74$ & $92.14$ & $0.330$ & $22.8$ & $12.4$ & $0.26$ & 0 & 0 & $0.028$ & $22.7$ & $24.9$ & $0.25$ & 0 & 0 \\ 
Sch. & & $0.4$ & $368$ & $124$ & $3.715$ & $49.4$ & $2.8$ & $0.93$ & $81.25$ & $84.61$ & $0.319$ & $26.5$ & $10.6$ & $0.27$ & 0 & 0 & $0.013$ & $26.4$ & $22.7$ & $0.26$ & 0 & 0 \\ 
Led. & & -- & -- & -- & $3.014$ & $49.6$ & -- & $0.61$ & $114.15$ & $69.54$ & $0.345$ & $19.1$ & -- & $0.25$ & 0 & 0 & $0.036$ & $19$ & -- & $0.25$ & 0 & 0 \\ 
Sch. & & -- & -- & -- & $3.025$ & $49.2$ & -- & $0.62$ & $113.02$ & $69.92$ & $0.348$ & $18.7$ & -- & $0.25$ & 0 & 0 & $0.031$ & $18.6$ & -- & $0.25$ & 0 & 0 \\ \hline
Led. & $60$ & $0.4$ & $346$ & $149$ & $4.393$ & $39.1$ & $6.7$ & $0.78$ & $69.84$ & $37.89$ & $0.361$ & $16.9$ & $30.4$ & $0.27$ & 0 & 0 & $0.047$ & $16.8$ & $52.1$ & $0.26$ & 0 & 0 \\ 
Sch. & & $0.4$ & $346$ & $138$ & $4.466$ & $38.5$ & $4.3$ & $0.82$ & $78.76$ & $49.34$ & $0.355$ & $18.1$ & $31.2$ & $0.28$ & 0 & 0 & $0.037$ & $18$ & $24.3$ & $0.28$ & 0 & 0 \\ 
Led. & & -- & -- & -- & $3.521$ & $36.4$ & -- & $0.48$ & $147.38$ & $59.45$ & $0.383$ & $14$ & -- & $0.27$ & 0 & 0 & $0.083$ & $13.9$ & -- & $0.27$ & 0 & 0 \\ 
Sch. & & -- & -- & -- & $3.531$ & $36.3$ & -- & $0.49$ & $145.23$ & $60.67$ & $0.396$ & $12.6$ & -- & $0.28$ & 0 & 0 & $0.114$ & $12.5$ & -- & $0.27$ & 0 & 0 \\ \hline
Led. & $40$ & $0.4$ & $314$ & $168$ & $5.769$ & $31.8$ & $4.3$ & $0.56$ & $13.40$ & $4.61$ & $0.417$ & $15.5$ & $62.9$ & $0.23$ & 0 & $0.001$ & $0.061$ & $15.4$ & 0 & $0.22$ & 0 & 0 \\ 
Sch. & & $0.4$ & $314$& $166$ & $5.699$ & $32$ & $19.4$ & $0.55$ & $12.73$ & $4.31$ & $0.423$ & $12.4$ & $50.5$ & $0.28$ & 0 & 0 & $0.123$ & $12.3$ & $56.8$ & $0.28$ & 0 & 0 \\ 
Led. & & -- & -- & -- & $4.433$ & $36.5$ & -- & $0.27$ & $0.29$ & $0.12$ & $0.472$ & $13$ & -- & $0.16$ & 0 & 0 & $0.087$ & $12.9$ & -- & $0.16$ & 0 & 0 \\ 
Sch. & & -- & -- & -- & $4.441$ & $36.5$ & -- & $0.27$ & $0.29$ & $0.12$ & $0.472$ & $13.9$ & -- & $0.932$ & $140.20$ & $70.80$ & $0.069$ & $13.7$ & -- & $0.99$ & $85.60$ & $88.30$ \\ \hline
Led. & $32$ & $0.4$ & $305$ & $188$ & $6.701$ & $28.1$ & $15.9$ & $0.45$ & $6.07$ & $1.89$ & $0.484$ & $14.3$ & 0 & $0.70$ & $159.09$ & $16.49$ & $0.117$ & $14$ & 0 & $0.94$ & $97.23$ & $62.10$ \\ 
Sch. & & $0.4$ & $306$ & $187$ & $6.640$ & $28.2$ & $12.7$ & $0.44$ & $5.84$ & $1.81$ & $0.521$ & $10.2$ & $52.8$ & $0.29$ & $0.01$ & $0.01$ & $0.213$ & $10.1$ & 0 & $0.28$ & $0.01$ & $0.01$ \\ 
Led. & & -- & -- & -- & $5.202$ & $30.1$ & -- & $0.27$ & $0.29$ & $0.12$ & $0.536$ & $11.5$ & -- & $0.99$ & $71.25$ & $90.27$ & $0.135$ & $11.3$ & -- & $0.99$ & $51.56$ & $90.04$ \\ 
Sch. & & -- & -- & -- & $5.209$ & $30.1$ & -- & $0.27$ & $0.29$ & $0.12$ & $0.542$ & $11.2$ & -- & $0.63$ & $145.69$ & $32.95$ & $0.186$ & $10.9$ & -- & $0.99$ & $56.82$ & $77.97$ \\ \hline
Led. & $25$ & $0.4$ & $291$ & $208$ & $7.895$ & $23.6$ & $80.9$ & $0.34$ & $3.14$ & $0.86$ & $0.60$ & $10.4$ & $2$ & $0.64$ & $168.65$ & $5.19$ & & & & & & \\ 
Sch. & & $0.4$ & $295$ & $209$ & $7.903$ & $23.6$ & $82.2$ & $0.34$ & $3.13$ & $0.86$ & $0.618$ & $9.9$ & $0.9$ & $0.82$ & $138.02$ & $14.18$ & $0.269$ & $9.7$ & 0 & $0.93$ & $100.50$ & $25.12$ \\ 
Led. & & -- & -- & -- & $6.324$ & $24.2$ & -- & $0.27$ & $0.29$ & $0.12$ & $0.624$ & $9.5$ & -- & $0.70$ & $143.82$ & $59.48$ & 0.296 & $9.3$ & -- & $0.84$ & $102.67$ & $68.93$ \\ 
Sch. & & -- & -- & -- & $6.312$ & $24.2$ & -- & $0.27$ & $0.29$ & $0.12$ & $0.692$ & $8.8$ & -- & $0.54$ & $195.70$ & $6.69$ & $0.433$ & $8.3$ & -- & $0.83$ & $119.92$ & $72.30$ \\ \hline
Led. & $20$ & $0.4$ & $279$ & $216$ & $9.536$ & $19.5$ & $173$ & $0.30$ & $2.35$ & $0.59$ & $0.780$ & $7.8$ & $0$ & $0.41$ & $143.10$ & $1.45$ & $0.927$ & $7.3$ & 0 & $0.49$ & $151.00$ & $6.60$ \\ 
Sch. & & $0.4$ & $274$ & $217$ & $9.507$ & $19.5$ & $175$ & $0.30$ & $2.42$ & $0.60$ & $0.85$ & $7.6$ & $0$ & $0.60$ & $106.34$ & $3.73$ & $0.866$ & $7.2$ & 0 & $0.75$ & $127.30$ & $15.88$ \\ 
Led. & & -- & -- & -- & $7.822$ & $19.7$ & -- & $0.27$ & $0.29$ & $0.12$ & $0.854$ & $7.3$ & -- & $0.63$ & $151.76$ & $53.99$ & $0.911$ & $7$ & -- & $0.71$ & $128.11$ & $64.66$ \\ 
Sch. & & -- & -- & -- & $7.743$ & $19.7$ & -- & $0.27$ & $0.29$ & $0.12$ & $0.873$ & $9$ & -- & $0.50$ & $57.86$ & $3.89$ & $1.219$ & $8.6$ & -- & $0.51$ & $81.52$ & $4.39$ \\ \hline
Led. & $15$ & $0.4$ & $259$ & $202$ & $13.610$ & $14.7$ & $145$ & $0.29$ & $2.37$ & $0.54$ & $1.134$ & $12.5$ & $0.3$ & $0.36$ & $6.11$ & $0.99$ & $2.944$ & $12.4$ & $0.2$ & $0.36$ & $6.36$ & $1.00$ \\ 
Sch. & & $0.4$ & $271$ & $201$ & $13.449$ & $14.7$ & $160$ & $0.29$ & $2.34$ & $0.53$ & $1.450$ & $11.2$ & $0.1$ & $0.41$ & $8.31$ & $1.24$ & $1.374$ & $11.1$ & $0$ & $0.43$ & $9.52$ & $1.35$ \\ 
Led. & & -- & -- & -- & $11.153$ & $14.8$ & -- & $0.27$ & $0.29$ & $0.12$ & $1.221$ & $13.1$ & -- & $0.28$ & $1.68$ & $0.45$ & $3.684$ & $13$ & -- & $0.29$ & $1.84$ & $0.49$ \\ 
Sch. & & -- & -- & -- & $11.021$ & $14.8$ & -- & $0.27$ & $0.29$ & $0.12$ & $1.314$ & $13.3$ & -- & $0.30$ & $2.04$ & $0.51$ & $4.79$ & $13.2$ & -- & $0.34$ & $3.07$ & $0.74$ \\ \hline
Led. & $12$ & $0.4$ & $248$ & $196$ & $18.711$ & $11.9$ & $173$& $0.28$ & $1.93$ & $0.45$ & $1.723$ & $11.2$ & $0.5$ & $0.34$ & $6.00$ & $0.89$ & $6.434$ & $11.1$ & $0.3$ & $0.34$ & $6.41$ & $0.91$ \\ 
Sch. & & $0.4$ & $262$ & $198$ & $18.373$ & $11.9$ & $193$ & $0.28$ & $1.89$ & $0.45$ & $2.097$ & $10.3$ & $0.4$ & $0.33$ & $5.33$ & $0.83$ & $3.566$ & $10.2$ & $0.3$ & $0.35$ & $6.27$ & $0.92$ \\ 
Led. & & -- & -- & -- & $15.395$ & $11.9$ & -- & $0.27$ & $0.29$ & $0.12$ & $1.862$ & $11.1$ & -- & $0.28$ & $1.58$ & $0.42$ & $7.60$ & $11.1$ & -- & $0.29$ & $1.74$ & $0.46$ \\
Sch. & & -- & -- & -- & $15.336$ & $11.9$ & -- & $0.27$ & $0.29$ & $0.12$ & $2.123$ & $11.4$ & -- & $0.30$ & $1.84$ & $0.49$ & $7.927$ & $11.3$ & -- & $0.31$ & $1.98$ & $0.53$ \\ \hline
Led. & $9$ & $0.4$ & $230$ & $185$ & $31.638$ & $9$ & $192$ & $0.27$ & $1.34$ & $0.36$ & $3.369$ & $8.8$ & $1.6$ & $0.33$ & $5.44$ & $0.82$ & $1.794$ & $8.7$ & $0.9$ & $0.33$ & $5.73$ & $0.83$ \\ 
Sch. & & $0.4$ & $248$ & $188$ & $31.216$ & $9$ & $198$ & $0.27$ & $1.40$ & $0.37$ & $3.760$ & $8.6$ & $0.6$ & $0.33$ & $5.72$ & $0.83$ & $3.143$ & $8.5$ & $0.3$ & $0.34$ & $6.10$ & $0.86$ \\ 
Led. & & -- & -- & -- & $26.268$ & $9$ & -- & $0.27$ & $0.29$ & $0.12$ & $3.404$ & $8.7$ & -- & $0.27$ & $1.03$ & $0.27$ & & & & & & \\ 
Sch. & & -- & -- & -- & $26.274$ & $9$ & -- & $0.27$ & $0.29$ & $0.12$ & $3.490$ & $8.8$ & -- & $0.28$ & $1.61$ & $0.42$ & & & & & & \\ \hline
Led. & $7$ & $0.4$ & $219$ & $175$ & $51.514$ & $7$ & $184$ & $0.27$ & $0.94$ & $0.28$ & $6.505$ & $6.9$ & $2.1$ & $0.32$ & $4.57$ & $0.75$ & & & & & & \\ 
Sch. & & $0.4$ & $235$ & $178$ & $51.001$ & $7$ & $188$ & $0.27$ & $0.97$ & $0.29$ & $6.901$ & $6.9$ & $2$ & $0.32$ & $4.68$ & $0.76$ & & & & & & \\
Led. & & -- & -- & -- & $43.078$ & $7$ & -- & $0.27$ & $0.29$ & $0.12$ & $6.506$ & $6.9$ & -- & $0.27$ & $1.05$ & $0.28$ & & & & & & \\ 
Sch. & & -- & -- & -- & $41.745$ & $7$ & -- & $0.27$ & $0.29$ & $0.12$ & $6.914$ & $6.9$ & -- & $0.28$ & $1.52$ & $0.40$ & & & & & & \\ 
\hline 
\end{tabular}} 
\label{TabListModels} 
\end{table*} 

%20-40Msol
The mass domain between 20 and 40\,M$_\sun$ is a transition domain between the stars that evolve into the red supergiant stage and remain a red supergiant until the end of their lifetimes and the stars, for instance the 60\,M$_\sun$ models, that never evolve into a RSG, but become a luminous blue variable before evolving into a WR phase. In this transition mass domain, the star can end its evolution as a red, yellow, or blue supergiant, or become a Wolf-Rayet star (in the latter case after having been a red supergiant for a while). The tracks show a complex behaviour resulting from intricate effects involving convection, rotational mixing, and mass loss.\\

Below an initial mass M$_{\rm RSG}\sim15\,{\rm M}_\odot$, stars end their lives as red supergiants (see Sect.~\ref{RSG}). Stars initially between M$_{\rm RSG}$ and M$_{\rm RSG-WR}\sim20-25\,{\rm M}_\odot$ end their lives as a yellow or blue supergiants. For initial masses between M$_{\rm RSG-WR}$ and M$_{\rm WR}\sim40\,{\rm M}_\odot$, stars end their lives as Wolf-Rayet stars after a red supergiant stage. We note that this mass range of stars that experience both an RSG and a WR phase is expected to be narrow because at the moment, apart from Westerlund-1 \citep[see e.g.][we present isochrones of our models compared with observed data of massive evolved stars in Westerlund-1 in Sect.~\ref{comparison_observations}]{Clark2005,Crowther2006,Negueruela2010,Beasor2021}, there is no single-age stellar population in which both red supergiants and Wolf-Rayet stars are observed at the same time. Finally, stars with initial masses above M$_{\rm WR}$ enter the WR phase without previously becoming a RSG.

These limits appear to be more sensitive to rotation than to the convective criterion (at least for the initial rotation speed considered here). In general, rotation tends to decrease the limits M$_{\rm RSG-WR}$ and M$_{\rm WR}$ (see more below). Because M$_{\rm WR}$ is shifted to lower values than M$_{\rm RSG-WR}$ in rotating models, this implies that rotation suppresses or at least disfavours the production of WR stars with RSG progenitors (at least for a single-star evolution).

Mass loss by stellar winds (which increases when the luminosity increases and the effective temperature decreases) begins to become a dominant feature in this mass domain (it becomes an even more important feature for higher initial masses). As a numerical example, the 20\,M$_\odot$ models lose significant amounts of mass during the post-MS phase (see Table~\ref{TabListModels}), ending core He-burning with a total mass of about 7-9\,M$_\odot$ (for comparison, the 15\,M$_\odot$ models reach the end of He-burning at around 11-13\,M$_\odot$). The masses of the 25\,M$_\odot$ models at the end of He-burning are about 9-10\,M$_\odot$ (this is higher than for the 20\,M$_\odot$ models, but more mass has been lost by the 25\,M$_\odot$ models). The 32 and 40\,M$_\odot$ models experience even stronger mass loss (losing up to 22 and 28\,M$_\odot$ for the 32 and 40\,M$_\odot$ models, respectively). They all end core He-burning at high ($\log{T_{\rm eff}~\rm{[K]}}>4.4$) effective temperatures, and some of them lose so much mass that they become Wolf-Rayet stars. 

While all the stars including and above 20\,M$_\odot$ lose large amounts of mass, the mass of the core remains an increasing function of the initial mass. In other words, the 20\,M$_\odot$ models reach the end of He-burning with a lower total mass than the 15\,M$_\odot$ models, but their CO cores are more massive (3.65-4.44\,M$_\odot$ compared with 2.11-2.66\,M$_\odot$). Furthermore, the models between 20 and 40\,M$_\odot$ experience much stronger surface nitrogen enrichment than those below 20\,M$_\odot$. 

% WR PHASES DURATIONS
 \begin{figure*}[h!]
 \centering
 \includegraphics[scale=1]{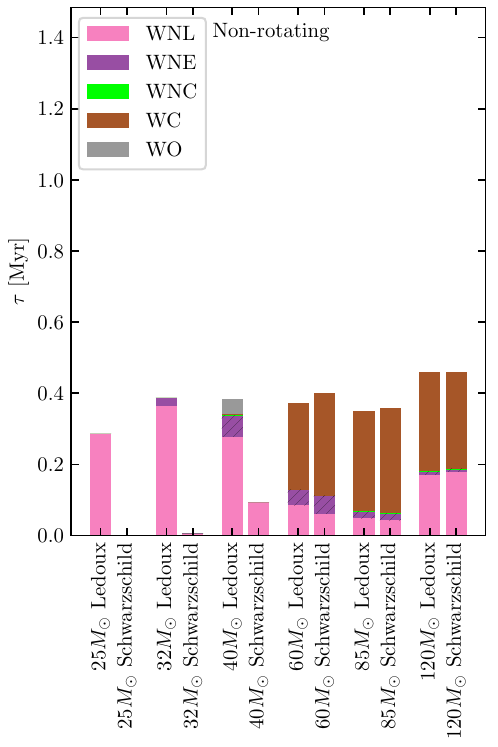}\includegraphics[scale=1]{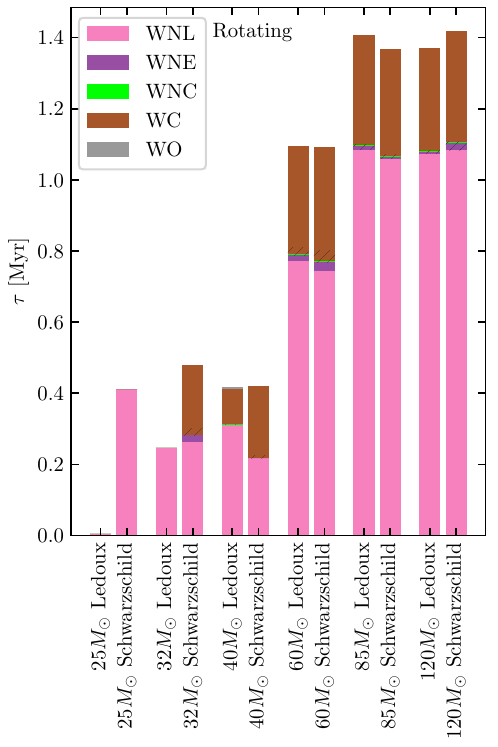}
 \caption{Bar plot showing the durations of the different Wolf-Rayet phases for masses at which at least one of the four (Ledoux, Schwarzschild, rotating, and non-rotating) models has a Wolf-Rayet phase. Left panel: Non-rotating models. Right panel: Rotating models.}
 \label{WR_durations}
 \end{figure*}

%60-120Msol
The mass domain above this transition mass range is dominated by mass loss. Not many differences are therefore visible between Schwarzschild and Ledoux models. 
The most noticeable difference occurs for the 120\,$M_\sun$ models, as discussed in Sect.~\ref{internal_structure}.

Figure~\ref{WR_durations} shows the durations of the different Wolf-Rayet phases (stacked in chronological order) for the masses where at least one of the four models reaches a WR phase. A first striking difference appears between the rotating and non-rotating models. Rotation produces longer WR phases. This is in line with previous studies \citep{MeM2003,Georgy2012}. The time spent as WR stars is longer for rotating than for non-rotating models because rotation drives additional mass loss and mixing. More mixing implies that less mass loss is required to uncover hydrogen-poor layers. Rotation also favours the WNL phase, decreases WNE and WC durations, and prevents the WO phase. Interestingly, rotation increases the WNC-phase duration.

Above and including 60\,M$_\sun$ the Schwarzschild and Ledoux models spend similar times (within 6\%) as Wolf-Rayet stars because mass loss is the dominating effect here. 
This agrees with the fact that the differences between Ledoux and Schwarzschild models are mostly due to the ICZ. They do not appear in the most massive stars due to the strong mass loss. As a result, the choice of criterion for convection intervenes as a second-order effect.

For the mass range between 25 and 40\,M$_\odot$, the differences induced by changing from the Ledoux to Schwarzschild criterion are more marked.
Without rotation, the Ledoux criterion produces Wolf-Rayet stars for lower initial masses than the Schwarzschild criterion. These Ledoux models spend a longer time of their core He-burning phase as an RSG than the Schwarzschild models. As a result, mass loss is stronger for the former, and they reach the WR phase earlier (as a result, the WR phase lasts longer).

\section{Final properties of the models}
\label{final_properties}

\begin{table}[!]
\caption{Properties of all the stars at the last computed model, remnant types and masses, and averaged mass fractions of different elements above the remnant mass.}
\centering
\scalebox{.6}{
\begin{threeparttable}
\begin{tabular}{|ccc|ccc|ccccccc|}
\hline\hline
Crit. & M$_\text{ini}$ & ${\frac{\upsilon_\text{ini}}{\upsilon_\text{crit}}}$ & M$_\text{fin}$ & $\log{T_\text{eff, fin}}$ & $\log{L_\text{fin}}$ & \multicolumn{2}{c}{Remnant} & $^1$H & $^4$He & $^{12}$C & $^{16}$O & $\mathcal{Z}$\tnote{$\dagger$}\\ 
& $M_{\sun}$ && M$_\odot$ & K & $L_\odot$ & Type & M$_\odot$ & \multicolumn{5}{c|}{mass frac.} \\ 
\hline
Led. & 120 & 0.4 & 30.8 & 5.4 & 6.2 & BH & 30.8 & 0.00 & 0.11 & 0.343 & 0.494 & 0.049 \\
Sch. & 120 & 0.4 & 19.0 & 5.4 & 5.9 & BH & 19.0 & 0.00 & 0.12 & 0.395 & 0.418 & 0.064 \\
Led. & 120 & -- & 28.6 & 5.3 & 6.2 & BH & 28.6 & 0.00 & 0.12 & 0.358 & 0.469 & 0.051 \\
Sch. & 120 & -- & 30.9 & 5.3 & 6.2 & BH & 30.9 & 0.00 & 0.12 & 0.346 & 0.469 & 0.061 \\ \hline
Led. & 85 & 0.4 & 22.7 & 5.4 & 6.0 & BH & 22.7 & 0.00 & 0.11 & 0.374 & 0.430 & 0.082 \\
Sch. & 85 & 0.4 & 26.4 & 5.4 & 6.1 & BH & 26.4 & 0.00 & 0.12 & 0.367 & 0.451 & 0.058 \\
Led. & 85 & -- & 19.0 & 5.4 & 6.0 & BH & 19.0 & 0.00 & 0.12 & 0.389 & 0.424 & 0.066 \\
Sch. & 85 & -- & 18.6 & 5.4 & 6.0 & BH & 18.6 & 0.00 & 0.12 & 0.391 & 0.426 & 0.065 \\ \hline
Led. & 60 & 0.4 & 16.8 & 5.4 & 5.9 & BH & 16.8 & 0.00 & 0.12 & 0.394 & 0.428 & 0.060 \\
Sch. & 60 & 0.4 & 18.0 & 5.4 & 5.9 & BH & 18.0 & 0.00 & 0.12 & 0.396 & 0.418 & 0.064 \\
Led. & 60 & -- & 13.9 & 5.4 & 5.8 & BH & 13.9 & 0.00 & 0.13 & 0.411 & 0.406 & 0.057 \\
Sch. & 60 & -- & 12.5 & 5.3 & 5.7 & BH & 12.5 & 0.00 & 0.12 & 0.413 & 0.401 & 0.061 \\ \hline
Led. & 40 & 0.4 & 15.4 & 5.4 & 5.8 & BH & 15.4 & 0.00 & 0.10 & 0.389 & 0.458 & 0.051 \\
Sch. & 40 & 0.4 & 12.3 & 5.3 & 5.6 & BH & 12.3 & 0.00 & 0.12 & 0.411 & 0.420 & 0.045 \\
Led. & 40 & -- & 12.9 & 5.3 & 5.7 & BH & 12.9 & 0.00 & 0.09 & 0.370 & 0.478 & 0.061 \\
Sch. & 40 & -- & 13.7 & 4.6 & 5.7 & BH & 13.7 & 0.00 & 0.56 & 0.165 & 0.204 & 0.073 \\ \hline
Led. & 32 & 0.4 & 14.0 & 4.5 & 5.7 & BH & 14.0 & 0.00 & 0.53 & 0.207 & 0.225 & 0.031 \\
Sch. & 32 & 0.4 & 10.1 & 5.3 & 5.5 & BH & 10.1 & 0.00 & 0.12 & 0.403 & 0.441 & 0.032 \\
Led. & 32 & -- & 11.3 & 4.6 & 5.6 & BH & 11.3 & 0.00 & 0.58 & 0.150 & 0.210 & 0.058 \\
Sch. & 32 & -- & 10.9 & 4.6 & 5.6 & BH & 10.9 & 0.00 & 0.71 & 0.112 & 0.119 & 0.058 \\ \hline
Led. & 25 & 0.4 & 10.4 & 4.3 & 5.4 & BH & 10.4 & 0.02 & 0.58 & 0.183 & 0.188 & 0.030 \\
Sch. & 25 & 0.4 & 9.7 & 4.4 & 5.5 & NS & 1.57 & 0.00 & 0.20 & 0.209 & 0.525 & 0.070 \\
Led. & 25 & -- & 9.3 & 4.4 & 5.5 & BH & 9.3 & 0.01 & 0.74 & 0.106 & 0.105 & 0.042 \\
Sch. & 25 & -- & 8.3 & 4.4 & 5.4 & NS & 1.55 & 0.00 & 0.33 & 0.171 & 0.417 & 0.081 \\ \hline
Led. & 20 & 0.4 & 7.3 & 3.8 & 5.3 & NS & 1.65 & 0.00 & 0.30 & 0.219 & 0.454 & 0.028 \\
Sch. & 20 & 0.4 & 7.2 & 4.3 & 5.3 & NS & 1.65 & 0.00 & 0.29 & 0.165 & 0.427 & 0.113 \\
Led. & 20 & -- & 7.0 & 4.0 & 5.2 & NS & 1.51 & 0.01 & 0.37 & 0.134 & 0.351 & 0.141 \\
Sch. & 20 & -- & 8.6 & 3.6 & 5.2 & NS & 1.77 & 0.17 & 0.48 & 0.072 & 0.174 & 0.105 \\ \hline
Led. & 15 & 0.4 & 12.4 & 3.6 & 5.1 & NS & 1.55 & 0.42 & 0.40 & 0.047 & 0.100 & 0.036 \\
Sch. & 15 & 0.4 & 11.1 & 3.6 & 5.1 & BH & 11.1 & 0.39 & 0.45 & 0.045 & 0.094 & 0.019 \\
Led. & 15 & -- & 13.0 & 3.6 & 4.9 & NS & 1.4 & 0.51 & 0.36 & 0.032 & 0.065 & 0.037 \\
Sch. & 15 & -- & 13.2 & 3.6 & 4.9 & NS & 1.4 & 0.49 & 0.41 & 0.025 & 0.045 & 0.033 \\ \hline
Led. & 12 & 0.4 & 11.1 & 3.6 & 4.8 & NS & 1.4 & 0.49 & 0.42 & 0.025 & 0.051 & 0.021 \\
Sch. & 12 & 0.4 & 10.2 & 3.6 & 4.9 & NS & 1.4 & 0.46 & 0.38 & 0.041 & 0.102 & 0.026 \\
Led. & 12 & -- & 11.1 & 3.5 & 4.7 & NS & 1.4 & 0.56 & 0.37 & 0.018 & 0.030 & 0.017 \\
Sch. & 12 & -- & 11.3 & 3.6 & 4.7 & NS & 1.4 & 0.56 & 0.37 & 0.016 & 0.031 & 0.019 \\ \hline
Led. & 9 & 0.4 & 8.7 & 3.6 & 4.3 & WD & 0.93 & 0.52 & 0.39 & 0.016 & 0.039 & 0.031 \\
Sch. & 9 & 0.4 & 8.5 & 3.5 & 4.6 & WD & 1.18 & 0.48 & 0.41 & 0.024 & 0.050 & 0.03 \\
Led. & 9 & -- & 8.7 & 3.6 & 4.1 & WD & 0.82 & 0.58 & 0.36 & 0.022 & 0.026 & 0.011 \\
Sch. & 9 & -- & 8.8 & 3.6 & 3.9 & WD & 0.83 & 0.58 & 0.36 & 0.007 & 0.020 & 0.034 \\ \hline
Led. & 7 & 0.4 & 6.9 & 3.6 & 3.7 & WD & 0.60 & 0.54 & 0.38 & 0.026 & 0.048 & 0.013 \\
Sch. & 7 & 0.4 & 6.9 & 3.6 & 3.8 & WD & 0.61 & 0.54 & 0.36 & 0.029 & 0.058 & 0.013 \\
Led. & 7 & -- & 6.9 & 3.6 & 3.8 & WD & 0.54 & 0.59 & 0.35 & 0.027 & 0.029 & 0.011 \\
Sch. & 7 & -- & 6.9 & 3.6 & 3.5 & WD & 0.57 & 0.59 & 0.33 & 0.028 & 0.035 & 0.014 \\ \hline
\end{tabular}
\begin{tablenotes}
\item Notes: We took the stellar properties at the end of core C-burning for the rotating 9\,M$_\odot$ models and all models 12\,M$_\odot$ and above, at the end of core He-burning for the 7\,M$_\odot$ and non-rotating 9\,M$_\odot$ models. In the case of direct collapse to a black hole, we give the averaged mass fractions of elements above the mass of the CO core.
\item [$\dagger$] Usually, Z refers to metallicity: $Z=1-X-Y$, where $X$ and $Y$ are the mass fractions of $^1$H and $^4$He, respectively. Because we refer to metallicity excluding $^{12}$C and $^{16}$O, we used the calligraphic $\mathcal{Z}=1-X-Y-X(^{12}\text{C})-X(^{16}\text{O})$.
\end{tablenotes}
\end{threeparttable}}
\label{tabpreSN}
\end{table} 

\begin{figure*}
 \includegraphics[scale=.5]{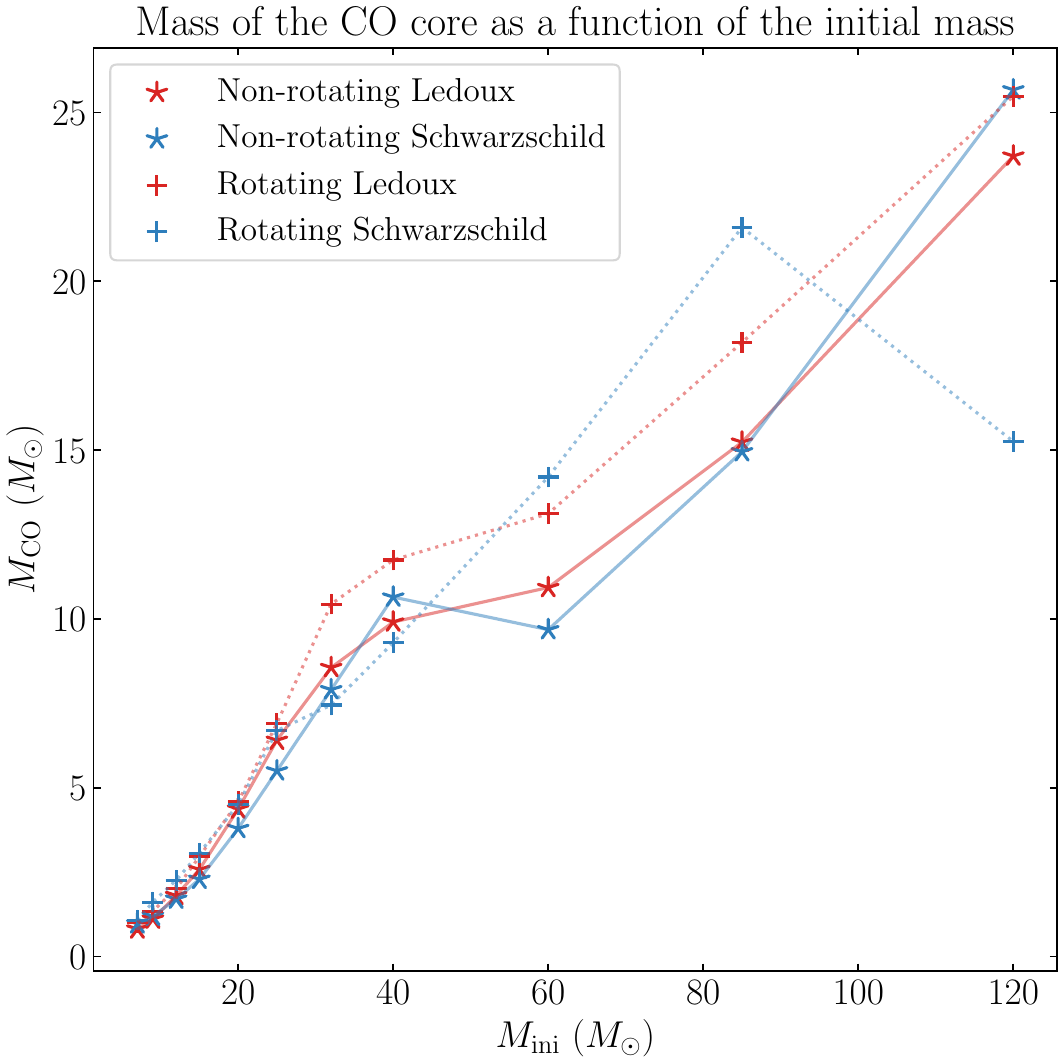}
 \includegraphics[scale=.5]{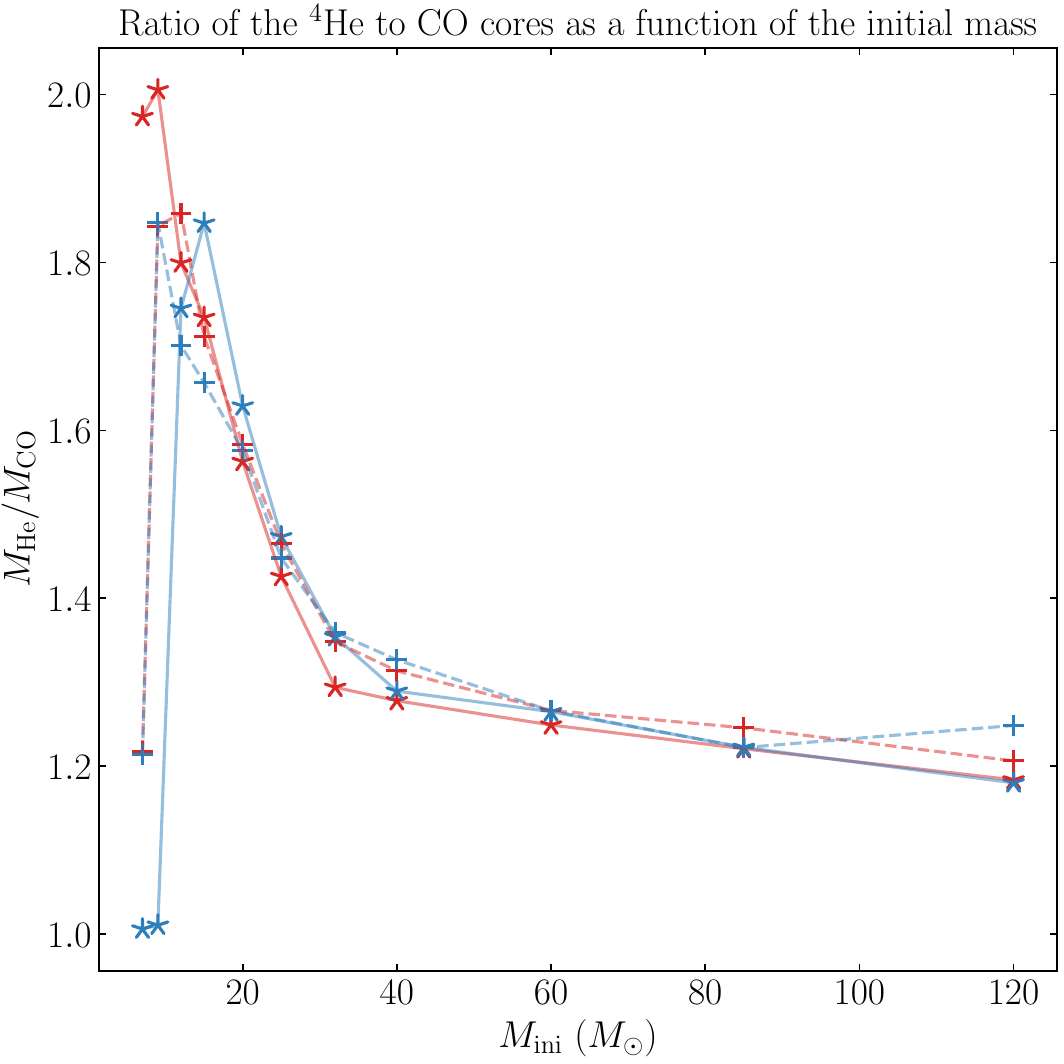} 
 \caption{Mass of the $^4\text{He}$ core (left panel) and mass ratio of the He and CO cores (right panel) at the last computed stage as a function of the initial mass for the different models. Stars (blue: Schwarzschild, and red: Ledoux) represent non-rotating models, and crosses (green: Schwarzschild, and orange: Ledoux) show rotating models.}
 \label{Mcores_vs_Mini}
\end{figure*}

\subsection{Core masses}

Figure~\ref{Mcores_vs_Mini} shows the CO-core masses and the mass ratios of the He and CO core as functions of the initial mass for the four sets of models at the end of their evolution. The masses of the cores are defined as in Table~\ref{internal} (here, we discuss their values at the last computed stage, however). The ratio of the mass of the He core and that of the CO core decreases from 2 to roughly 1.3 between 7 and 32\,M$_\odot$ and then remains constant around 1.2-1.3, indicating that the He-burning shell sits relatively farther out from the C-burning core in lower-mass models than in higher-mass models. 

The differences among models of the same mass in the left panel of Fig.~\ref{Mcores_vs_Mini} show that rotating models have a 10\% larger core on average (except for the rotating 32, 40, and 120\,M$_\odot$ Schwarzschild models) than non-rotating models. There is no clear trend in the differences between Schwarzschild and Ledoux models, however.
The impact of rotation is stronger than that of changing the convective criterion.

The strongest impact of changing from the Ledoux to Schwarzschild criterion appears for the 7 and 9\,M$_\odot$ stellar model. The ratio of He- to CO-core mass can vary from 1 to 2, depending on the criterion. This mass domain covers the transitions between stars that would produce white dwarfs and those that would produce neutron stars at the end of their lifetimes. This transition has been studied for instance by \citet{Siess2006, Siess2007}. This final fate depends on whether ignition of carbon occurs in degenerate, mildly degenerate, or non-degenerate conditions. A star whose central region is for a large part in mildly degenerate conditions can still succeed in igniting (off-centre) carbon, but this requires a more massive CO core than in non-degenerate conditions. It is therefore expected that a larger part of the central regions for models with the lowest ratio of the He- to the CO-core mass lies in the degenerate domain. Only models that ignite carbon in non-degenerate or very mildly degenerate conditions ({\it i.e.} the rotating 9 M$_\odot$ models) show high ratios of the He to CO core.

\subsection{Properties of the compact object progenitors}

%Present Table 4
We list integrated chemical abundances (in mass fractions) of $^1$H, $^4$He, $^{12}$C, and $^{16}$O in the models at the end of core C-burning (core He-burning for stars that do not reach the end of C-burning) in Table~\ref{tabpreSN}. This moment is close enough to the end of evolution that the abundances of the isotopes shown in the ejecta will not change dramatically \citep[see e.g. the Appendix Table in][]{Limongi2018}. Furthermore, explosive nucleosynthesis during the supernova is not expected to significantly affect the yields of these four elements. We computed the remnant types and masses using the models from \citet{Patton2020}, whose predictions concerning the remnant type (neutron star or black hole) were based on the `Ertl criterion' \citep{Ertl2016,Ertl2020}. For each star, we took the mass of the CO core $M_{\rm CO}$ (defined, like in Table~\ref{internal}, as the region within which the mass fraction of helium $Y<10^{-2}$) and the mass fraction of $^{12}$C in the core $X_{\rm ^{12}C}$ at the end of helium-burning. When $M_{\rm CO}<1.4\,M_\odot$, then we considered the remnant to be a white dwarf with a mass of $M_{\rm WD}=M_{\rm CO}$. When $1.4\,M_\odot<M_{\rm CO}<2.5\,M_\odot$, the remnant is a neutron star with a mass of $M_{\rm NS}=1.4\,M_\odot$. When $2.5\,M_\odot<M_{\rm CO}<10\,M_\odot$, we consulted the top panel of Fig.~3 of \citet{Patton2020} to determine whether the star will explode or implode (without ejecting anything). If it explodes, then we estimated the baryonic\footnote{We do not provide the gravitational masses of the neutron stars. One can compute them using the equation of state of their choice. One example for a relation between baryonic and gravitational mass can be found in Eq.~(39) of \citet{Griffiths2022}. Typically the gravitational mass of a neutron star will be 10-20\% smaller than its baryonic mass.} mass of the resulting neutron star to be $M_{\rm NS} = M_4$, where $M_4$ is defined in \citet[we use their Table~1 which provides the values of $M_4$ for a range of $(M_{\rm CO},X_{\rm ^{12}C})$ at the end of core helium-burning]{Patton2020}. If it implodes, then the mass of the resulting black hole is $M_{\rm BH} =M_{\rm fin}$. Finally, if $10\,M_\odot<M_{\rm CO}<30\,M_\odot$, the outcome is a black hole, with $M_{\rm BH} = M_{\rm fin}$. This equality between the black hole mass and the final mass of the star may be an overestimation, but the underlying assumption is that the entire star collapses into the black hole and that no matter is ejected. Any value between $M_{\rm CO}$ and $M_{\rm fin}$ would be a reasonable estimate for $M_{\rm BH}$. The case $M_{\rm CO}>30\,M_\odot$ could yield a pulsational or regular pair-instability supernova, but this applies to none of our stars.
We then integrated the abundances of $^1$H, $^4$He, $^{12}$C, and $^{16}$O above the remnant (between its mass coordinate and the surface). If the remnant is a black hole, we list the integrated chemical abundances above the mass coordinate of the CO core. We grouped all other elements into the quantity $\mathcal{Z}$ (usually, the metallicity $Z$ includes $^{12}$C and $^{16}$O, and therefore, we used the calligraphic $\mathcal{Z}= 1-X-Y - X(^{12}\text{C}) - X(^{16}\text{O})$). We also provide the mass, effective temperature, and luminosity of the final models. 

%Remnant types
Using the criteria defined above on $M_{\rm CO}$, as well as the grid of CO cores with varying $^{12}$C core mass fractions computed by \citet{Patton2020}, we can predict the type of compact object that would remain after the end of stellar evolution. All of the 7 and 9\,M$_\odot$ models would yield white dwarfs (WD in Table~\ref{tabpreSN}). Of the 12, 15, and 20\,M$_\odot$ models, the rotating 15\,M$_\odot$ Schwarzschild model would become a black hole, while all the others would become neutron stars. In the 25\,M$_\odot$ stars, the Ledoux models become black holes and the Schwarzschild models become neutron stars. All stars 32\,M$_\odot$ and above become black holes.

%PreSN yields
The composition of the envelopes shows that, as expected for stars that reach the WR phase, the models above 20\,M$_\odot$ do not have any $^1$H at the end of their evolution. More generally, the integrated abundances of both $^1$H and $^4$He decrease with increasing initial mass.

Conversely, the quantity of metals (including $^{12}$C and $^{16}$O) in the envelope is an increasing function of the initial mass. We find no striking consistent difference between the Schwarzschild and Ledoux models. The differences mostly concern the integrated $^{12}$C and $^{16}$O abundances: in some cases, one model produces more $^{12}$C and less $^{16}$O than its counterpart, but there is no clear-cut effect.

The case of the 25\,M$_\odot$ models is interesting because the two Ledoux models are predicted to become black holes and the Schwarzschild models to become neutron stars. There is a striking difference in the compositions of their envelopes: the former are much richer in $^4$He ($Y_{\rm env}$ of 0.58 and 0.74 against 0.20 and 0.33 for the Schwarzschild models). The latter, conversely, are much richer in $^{12}$C and $^{16}$O (with the most marked differences being in the $^{16}$O abundance: $X(^{16}\rm{O})_{\rm env}$ of 0.525 and 0.417 against 0.188 and 0.105 for the Ledoux models).

In the 32 and 40\,M$_\odot$ models, the rotating 32\,M$_\odot$ and non-rotating 40\,M$_\odot$ Schwarzschild models present very different $^{12}$C and $^{16}$O abundances from the other models of the same mass. The former is much richer in $^{12}$C and $^{16}$O (and thus poorer in $^4$He) than the other 32\,M$_\odot$ models; the latter is much poorer in $^{12}$C and $^{16}$O (and thus richer in $^4$He) than the other 40\,M$_\odot$ models.\\

%Expected type of SN 
From the $^1$H and $^4$He content of the models at the last computed stage, we can infer the type of core-collapse supernova that is expected to occur for all the exploding models. All the models of 12 and 15\,M$_\odot$ contain significant amounts of $^1$H (about 50\% of the envelope mass), and therefore, we expect type II SNe from them (except for the rotating Schwarzschild 15\,M$_\odot$ model, which we expect to directly collapse into a black hole). Of the 20\,M$_\odot$ stars, only the non-rotating Schwarzschild model is expected to lead to a type II SN because it still contains a non-negligible amount of $^1$H. We expect the other 20\,M$_\odot$ models to die in type Ib SNe because they have no $^1$H left, but still contain $^4$He. The same holds for the 25\,M$_\odot$ Schwarzschild models, for which we expect an explosion.

Finally, we theoretically predict the remnant masses. The most massive neutron star comes from the non-rotating 20\,M$_\odot$ Schwarzschild model, with a baryonic mass $M_{\rm NS}=1.77M_\odot$. The least massive black hole results from the non-rotating 25\,M$_\odot$ Ledoux model, with $M_{\rm BH}=9.3 {\rm M}_\odot$. Acknowledging that taking $M_{\rm BH} = M_{\rm fin}$ may be an overestimation of the black hole mass, we can estimate the absolute minimum predicted black hole mass from that of the CO cores (taking $M_{\rm BH} = M_{\rm CO}$). In this case, the 15\,M$_\odot$ rotating Schwarzschild model yields the lowest-mass black hole, with $M_{\rm BH}=2.68 {\rm M}_\odot$.

\section{Stellar population synthesis}
\label{Populations}
\subsection{The population synthesis models}

The population synthesis results we present in this work were computed using the \textsc{Syclist} code \citep{Georgy2014b}, using the stellar models from \citet[computed with the Schwarzschild criterion]{Ekstrom2012}, and from \citet{Saio2013a} and \citet[20 and 25\,M$_\odot$ models computed with the Ledoux criterion]{Georgy2014} as input. We added all the other Ledoux models that we computed for the current paper in order to cover masses from 7 to 120\,M$_\odot$. \textsc{Syclist} generates a stellar population with initial masses sampled according to the Salpeter IMF, and it saves the state of the population at each requested time step. To mimic continuous star formation rates, we summed the populations at each time step, and we then counted stars and classified them into subtypes according to some of their properties. The subtypes are defined in Table~\ref{TabSubtypes}.

We generated two types of clusters. One type consisted of a burst of star formation at $t=0$, and the other type had continuous star formation for 60\,Myr. The second type allowed us to simulate a stationary state and derive equilibrium relative abundances of stellar subtypes. We also generated isochrones of our populations to compare them to the observed populations of evolved massive stars in Westerlund-1.

\subsection{Case of a burst of star formation}
\label{Populations-burst}
\begin{figure*}[h!]
 \centering
 \includegraphics{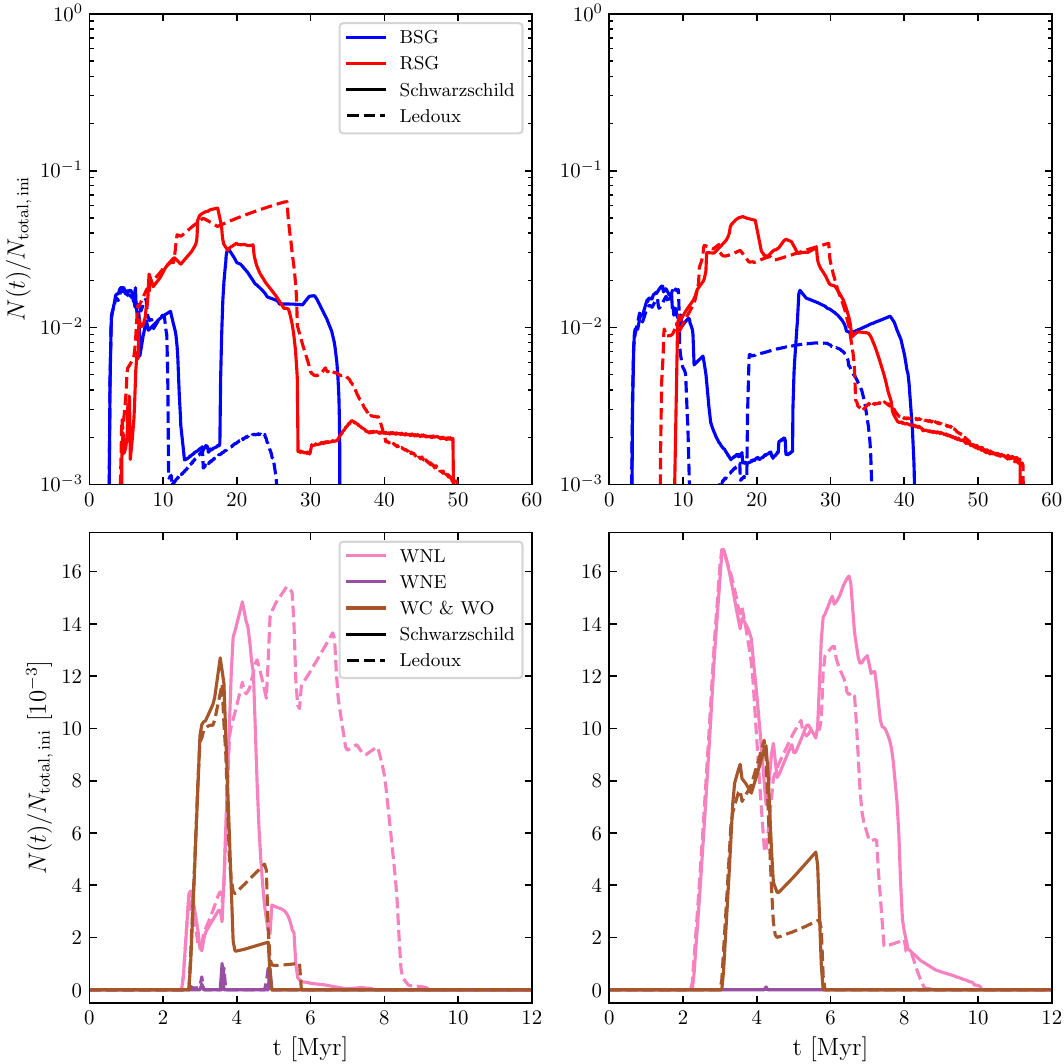}
 \caption{Number of stars of certain types alive at each time, normalised by the total initial number of stars generated in a synthetic cluster for the case of a burst of star formation at t=0. The solid lines represent stars computed with the Schwarzschild criterion, and the dashed lines show the same for the Ledoux criterion. Top panels: Blue and red supergiants. The vertical axis is logarithmic, and the time axis extends to 60\,Myr. Bottom panels: Wolf-Rayet stars. The vertical axis is linear and in units of $10^{-3}$, and the time axis extends to 12\,Myr (there are no WR stars later on). Left panels: Non-rotating models. Right panels: Rotating models.}
 \label{clusterdensity}
\end{figure*}

Figure~\ref{clusterdensity} shows the number of blue and red supergiants (top panels) and Wolf-Rayet stars (bottom panels) alive at each time in the case of a burst of star formation, normalised by the initial number of stars. For non-rotating stars, the Ledoux clusters produce far fewer blue and many more red supergiants than the Schwarzschild clusters. 
This is attributable to the different crossing durations of the Hertzsprung gap above 15\,M$_\odot$ (the Schwarzschild models start core He-burning as BSGs, the Ledoux models start as RSGs), and to a larger extent (because of the IMF slope) to the blue loop behaviour, which is present in Schwarzschild stars between 7 and 12\,M$_\odot$ but is absent in Ledoux models.

This is not so much the case for rotating stars because, as mentioned previously, rotation tends to mitigate the differences between the two sets of models, and the blue loop behaviour is much more similar in rotating models. The difference in these models occurs just before 20\,Myr, which is when the Ledoux 12\,M$_\odot$ model undergoes its blue loop while the Schwarzschild model does not.

Finally, we note that the non-rotating Wolf-Rayet populations (especially the WNL stars) are quite different, with Ledoux clusters clearly hosting more WR stars than Schwarzschild clusters. This is expcected because the WR stage is reached for non-rotating Ledoux stars of lower masses (down to 25\,M$_\odot$), and it lasts longer for these lower masses (see the left panel of Fig.~\ref{WR_durations}) than in the Schwarzschild models. WNE, WNC, and WO type stars are very rare in both types of clusters, and the differences in WC number densities are not significant.

\subsection{Stationary regime populations}
\label{Populatios-stationary}
% CLUSTER DENSITY MAP ALL STARS S0
 \begin{figure*}[h!]
 \centering
 \includegraphics{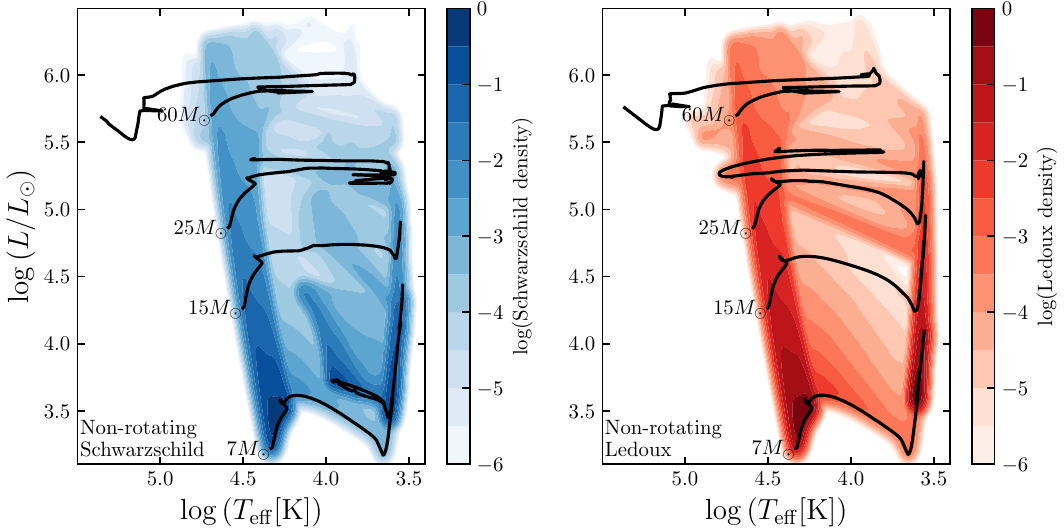}
 \caption{Density maps in the HRD of the Schwarzschild and Ledoux non-rotating populations in the case of continuous star formation (stationary regime). We add the tracks of the 7, 15, 25, and 60\,M$_\odot$ models. Left: Schwarzschild model. Right: Ledoux model.}
 \label{clusterS0contour}
 \end{figure*}

% CLUSTER DENSITY MAP ALL STARS S4
 \begin{figure*}[h!]
 \centering
 \includegraphics{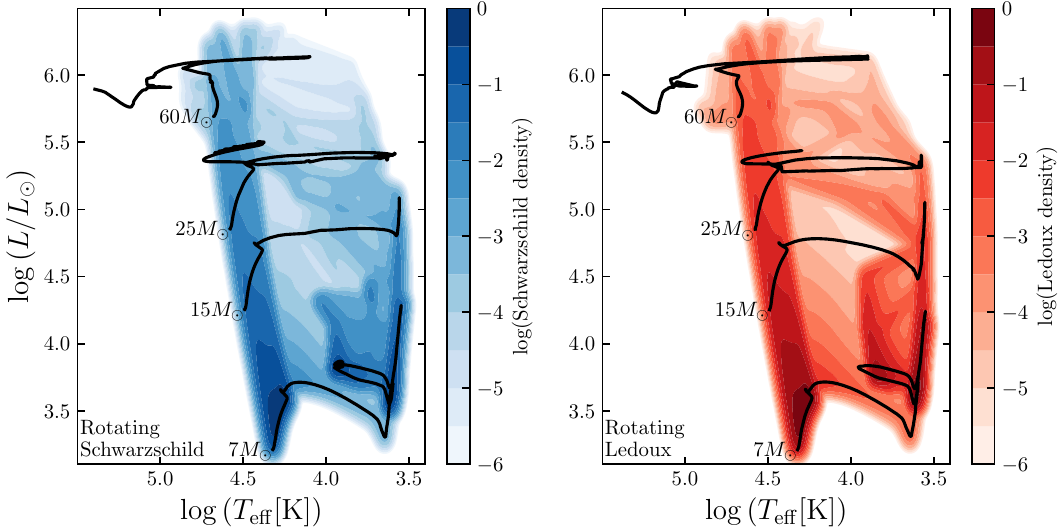}
 \caption{Density maps in the HRD of the Schwarzschild and Ledoux rotating populations in the case of continuous star formation (stationary regime). We add the tracks of the 7, 15, 25, and 60\,M$_\odot$ models. Left: Schwarzschild model. Right: Ledoux model.}
 \label{clusterS4contour}
 \end{figure*}

% CONTINUOUS STAR FORMING CLUSTERS' STELLAR ABUNDANCES TABLE
\begin{table}[h!]
\caption{Number ratios of different classes of stars (stationary regime population with continuous star formation for 60\,Myr).}
\centering
\scalebox{0.9}{\begin{tabular}{|lc|c|c|}
\hline \hline
Ratio & $\upsilon_{\rm ini}/\upsilon_{\rm crit}$ & Schwarzschild & Ledoux\\
\hline
Post-MS $/$ MS & $0$ & $0.152$ & $0.144$\\
 & $0.4$ & $0.134$ & $0.123$\\
BSG $/$ RSG & $0$ & $0.61$ & $0.13$\\
 & $0.4$ & $0.42$ & $0.31$\\
WR $/$ RSG & $0$ & $0.045$ & $0.070$\\
 & $0.4$ & $0.106$ & $0.096$\\
WNE $/$ WNL & $0$ & $6.1\times 10^{-3}$ & $2.8\times 10^{-3}$\\
 & $0.4$ & $8.5\times 10^{-5}$ & 0\\
WO $/$ WC & $0$ & $0$ & $0.082$\\
 & $0.4$ & $0$ & $0.014$\\
WNC $/$ WR & $0$ & $0.024$ & $0.013$\\
 & $0.4$ & $0.022$ & $0.017$\\
WN $/$ WR & $0$ & $0.60$ & $0.79$\\
 & $0.4$ & $0.81$ & $0.81$\\
WC $/$ WN & $0$ & $0.677$ & $0.243$\\
 & $0.4$ & $0.235$ & $0.232$\\
\hline 
\end{tabular}}
\label{TabClusterRatios} 
\end{table} 

We now discuss populations in the stationary regime. Table \ref{TabClusterRatios} shows the number ratios of different classes of stars in this equilibrium state, where we simulated populations with a constant continuous star formation episode lasting 60\,Myr. The ratios of post-MS to MS stars are very similar in Schwarzschild and Ledoux models: we find differences of 5\% (non-rotating) and 9\% (rotating), and the Schwarzschild post-MS to MS ratio is higher than the Ledoux ratio because Ledoux models have a slightly longer main sequence and shorter helium-burning phases. The ratios of BSGs to  RSGs are higher for the Schwarzschild populations regardless of the rotation scheme, although the difference is more marked for populations of non-rotating stars (the BSG to RSG ratio is almost five times higher for Schwarzschild than for Ledoux populations); this echoes what we mentioned in Sect.~\ref{Populations-burst}, and we attribute this difference to the different blue loop behaviour and to the durations of the Hertzsprung gap crossing.

The WR to RSG ratio is quite interesting in the non-rotating case. Ledoux populations produce more RSGs, but also more WR stars than Schwarzschild populations. The overproduction is higher for the WR stars, as $\rm{WR}/\rm{RSG}_{\rm Ledoux} >\rm{WR}/\rm{RSG}_{\rm Schwarzschild}$. Finally, while the populations produce similar numbers of WC stars (see the lower panels of Fig.~\ref{clusterdensity}), the ratio of WC to WN stars is three times higher for the non-rotating Schwarzschild population because the Ledoux population contains more WN(L) stars.

Figures~\ref{clusterS0contour} (non-rotating) and \ref{clusterS4contour} (rotating) show density maps in the HRD of populations generated with continuous star formation. The densities are higher in the lower parts of the diagram because more lower-mass stars are generated due to the Salpeter IMF we used. The comparison of populations both with and without rotation shows that the MS regions are the same for the Schwarzschild and Ledoux models. 

We note for the non-rotating populations that they contain a region of higher density that corresponds to the blue loops of the Schwarzschild models, but this region is absent from the Ledoux models. We also see differences in the supergiant stars above $\log{(L~[L_\odot])}\sim4.5$ because the 20 to 40\,M$_\odot$ models experience a blue and/or redward evolution in diverse parts of the HRD. 

For the rotating populations, the blue loop stars are present in both sets of models, but their high-density regions extend to lower surface temperatures for the Ledoux models ($\log{T_{\rm eff}~\rm{[K]}}\sim3.85$) than for the Schwarzschild models ($\log{T_{\rm eff}~\rm{[K]}}\sim3.95$). This would not be noticeable from the HRD tracks alone (the loops even extend farther bluewards for the rotating Ledoux stars), which means that the Ledoux models spend less time at the blue edge of the loop than the Schwarzschild models do. Fig.~\ref{clusterS4contour} shows that the higher-density region of the blue loops corresponds to the hottest point along the loop of 7\,M$_\odot$ Schwarzschild stars, but the equivalent point of the Ledoux-star loop extends to a higher temperature. The high-luminosity part of the blue-loop region reaches higher luminosities for the Ledoux models ($\log{(L~[L_\odot])}\sim4.6$) than for the Schwarzschild models ($\log{(L~[L_\odot])}\sim4.3$) because the threshold for blue loops is between 12 and 15\,M$_\odot$ for the former and between 9 and 12\,M$_\odot$ for the latter. We also see differences in the higher luminosity supergiants, which we explain in a similar way to those of the non-rotating models.

\section{Comparison with previous works and observations}
\label{comparisons}
\subsection{Comparison with previous theoretical works}
\label{comparison_previous}
While this paper presents the first grid that spans such a wide mass range in the comparison of Schwarzschild and Ledoux models at solar metallicity, with and without rotation, it is not the first to study the impact of various convective parameters on stellar evolution. In this section, we compare our conclusions with a few such previous theoretical works.

%Schootemeijer 2019
\citet{Schootemeijer2019} used MESA \citep{Paxton2011,Paxton2013,Paxton2015,Paxton2018} to compute models between 9 and 100\,M$_\odot$, with and without rotation, at SMC metallicity ($Z=0.002$). They used the Ledoux criterion and varied parameters related to the strength of overshooting and efficiency of semiconvective mixing. The models relevant to our comparison are those with overshooting parameter $\alpha_{\rm ov}=0.11$, and the most extreme values of the semiconvection parameter $\alpha_{\rm sc}$ = 0.01 and 300 (or 100 because they discussed more results related to that value). They found that very efficient semiconvection (large $\alpha_{\rm sc}$, Schwarzschild in the current paper) leads to more time being spent by stars as blue supergiants. For instance, these models start to burn helium in the core at a higher effective temperature than those with inefficient semiconvection. Fig.~2 (top panel for $\alpha_{\rm ov}=0.11$) of \citet{Schootemeijer2019} shows that blue loops occur in the lower-mass range for efficient semiconvection, but not when it is inefficient. All of these observations lead to a higher ratio of blue to red supergiants when semiconvective mixing is efficient. Qualitatively, this agrees very well with what we find. The specific values for the mass range of the blue loops, the effective temperatures at the beginning of core helium burning, and the ratio of blue to red supergiants differ from what we predict, but that is to be expected because their models were computed at SMC metallicity (which is almost an order of magnitude lower than ours).

%Kaiser 2020
\citet{Kaiser2020} also used MESA to compute non-rotating models of 15, 20, and 25\,M$_\odot$ at solar metallicity. They computed models with both the Schwarzschild and Ledoux criteria and used a free parameter $f_{\rm CBM}$, where CBM stands for convective boundary mixing, to vary the amount of additional mixing at the convective boundary (larger $f_{\rm CBM}$ corresponds to more mixing). This parameter intervenes in a diffusive model rather than the penetrative overshoot that we used in this study. They found that the main-sequence evolution is not affected by the choice of criterion for convective stability as long as CBM is included (in our case, as long as overshooting is included). They also discussed the intermediate convective zones and found that ICZs that are larger and remain for longer mean that stars continue to be blue until the ICZ recedes. Their stars computed with the Schwarzschild criterion have stronger (meaning larger and longer-living) ICZs than those computed with the Ledoux criterion. As a result, their Schwarzschild stars start core helium-burning as blue supergiants and the Ledoux stars start as red supergiants. Their 15\,M$_\odot$ Ledoux models also cross the HRD much faster after the main sequence, and this is correlated with a drop in surface luminosity. Overall however, their results are more affected by the value of $f_{\rm CBM}$ than by the choice of Schwarzschild and Ledoux criterion.

%Anders 2022
\citet{Anders2022} performed a 3D hydrodynamical simulation of a convective zone adjacent to a semiconvective (Ledoux-stable but Schwarzschild-unstable) region. They found that overshooting mixes the semiconvective region, increasing the size of the convective zone by a process they called entrainment. After a few thousand convective-overturn times, the Ledoux and Schwarzschild criteria predict the same convective boundary. The consequence is that as the title of \citet{Anders2022} states, `Schwarzschild and Ledoux are equivalent on evolutionary timescales'. This is the case during the main sequence (when the evolutionary timescale is much longer than the convective-overturn timescale). However, when the two timescales are of similar order, Ledoux predicts the instantaneous location of the boundary, whereas Schwarzschild provides its location in a stationary state. This agrees with what we find: that the largest differences are introduced during the post-MS expansion, which occurs on a rapid timescale.

\subsection{Comparison with observations}
\label{comparison_observations}
\begin{figure*}[h!]
    \centering
    \includegraphics{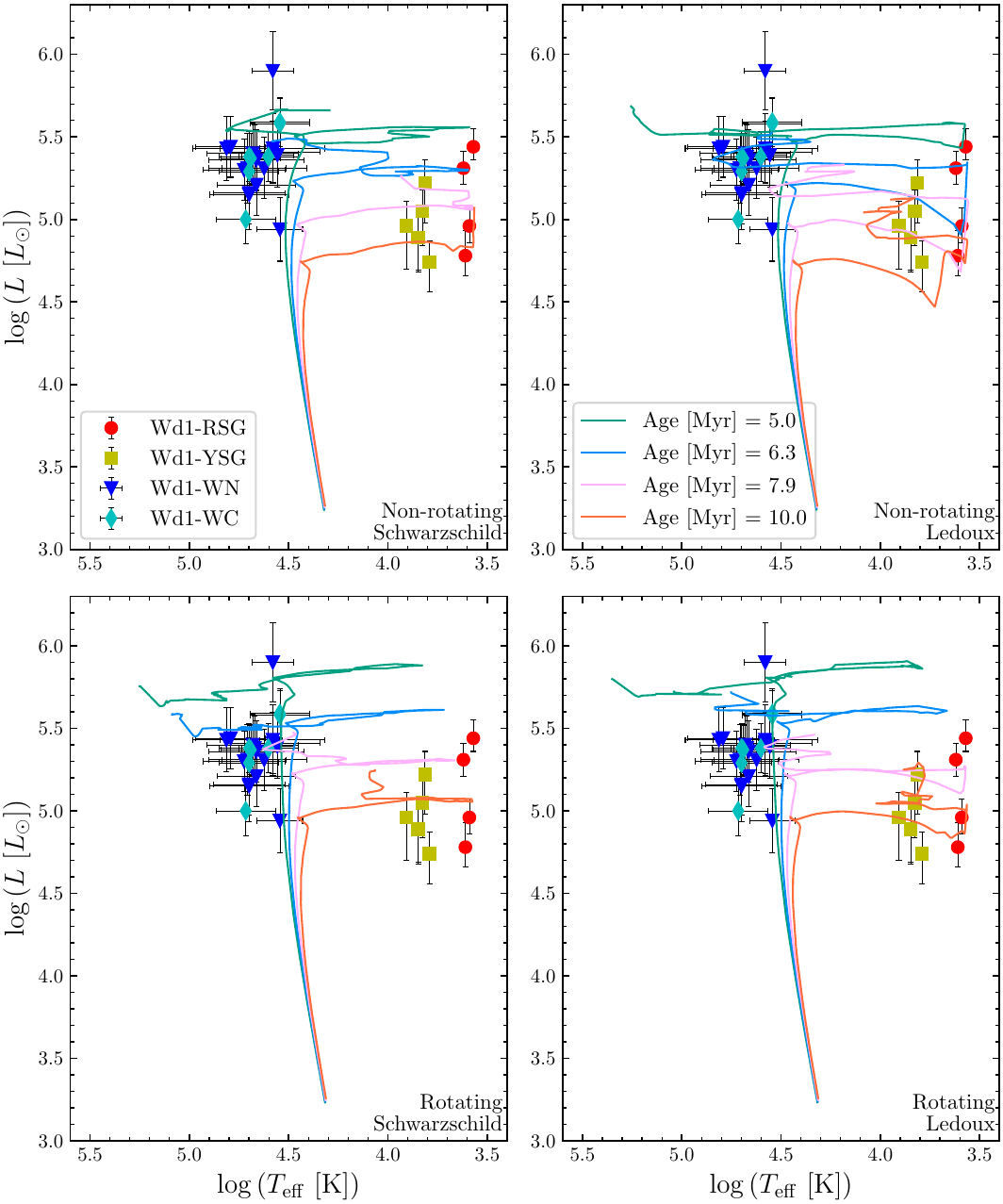}
    \caption{HRD showing isochrones at $\log{(\rm{Age~[yr]})}=6.7,\,6.8,\,6.9,\,{\rm and}~7.0$ (corresponding to ages of 5, 6.3, 7.9, and 10\,Myr) for our four sets of grids, compared with observations of evolved massive stars in the Westerlund-1 cluster. We use the data from \citet{GridZ020} for the observed stars. Left column: Schwarzschild models. Right column: Ledoux models. Upper row: Non-rotating models. Lower row: Rotating models.}
    \label{Wd1}
\end{figure*}
We briefly present here a comparison of our four sets of models with the evolved massive stars in the Westerlund-1 (Wd1) cluster. Wd1, at a distance of $\sim$4\,kpc \citep{Beasor2021}, is the best-studied young cluster in the galactic disk,  and it contains both RSG, YSG, and WR stars. Its metallicity is unknown because the gas associated with its formation has been dispersed. Based on its age, however, we may expect that its metallicity could be higher than that in our study (which is why \citet{GridZ020} used it as a comparison for their super-solar metallicity ($Z=0.020$) models). We nevertheless chose to also compare our models to Wd1 because it is the best-studied cluster containing both cool supergiants and Wolf-Rayet stars. We again used \textsc{SYCLIST} \citep{Georgy2014b}, this time, to generate isochrones for our four grids of models. We used the data from \citet{GridZ020} for the observed evolved stars in Wd1.

Figure~\ref{Wd1} shows isochrones for our four sets of grids and the observed data of evolved massive stars in Wd1. While no isochrone perfectly fits the observed distribution of stars, we find that the non-rotating Ledoux isochrone at an age of 6.3\,Myr best incorporates the presence of RSG and WR stars, as well as the spread in luminosity of the RSGs. The 10\,Myr Ledoux isochrones (both rotating and non-rotating) fit the observed YSGs. We recall the possibility that a cluster formation event lasted a few million years, typically between 10 and 6\,Myr ago, and that real stellar populations have varied rotation rates. Moreover, multiple star evolution and cluster dynamics will impact the observable properties of stars and induce a scatter in their luminosities, and this can explain some of the observed data.

\section{Conclusions and discussion}
\label{Conclusion}

In the above sections, we discussed the possible effect of changing the convective criterion on stellar model outputs with and without rotation. 
In many aspects, the effects appear to be relatively minor compared to those of rotation or mass-loss. We summarise the main results from this work below.
\begin{itemize}
% MS
\item The main-sequence phase is not much affected by the choice of the convection criterion.
% Convective zones
\item The Schwarzschild criterion does not necessarily lead to more extended convective zones. It does facilitate the formation of an extended intermediate convective zone at the end of the main sequence (especially for stars with initial masses between 15 and 32\,M$_\odot$), however, which gives rise to significant differences in the post-MS evolution of these stars.
% HRD crossing 
\item One such difference occurs during the first crossing of the HRD, where stars computed with the Schwarzschild criterion can start core He-burning as BSG, whereas the Ledoux stars start as RSG. This has important implications in the subsequent mass loss of the models. This in turn affects the ratio of blue to red supergiants as well as the way in which case B mass transfer occurs in close binary systems. Because the Schwarzschild criterion causes longer crossing times, it predicts more nitrogen-poor Cepheids (those undergoing their first crossing of the HRD) than the Ledoux criterion. Observations of nitrogen-poor Cepheids would definitely tip the scale in favour of the Schwarzschild criterion (or efficient semiconvection).
% Blue loops 
\item The occurrence of blue loops is affected by the choice of a Schwarzschild and Ledoux criterion: non-rotating 7-9\,M$_\odot$ Ledoux stars do not have them, but Schwarzschild stars do. This is probably due to different chemical profiles and helium abundances near and above the H-burning shell, and it also affects the ratio of blue to red supergiants.
% Cepheids
\item The duration of and the surface velocities (for rotating stars) during the Cepheid phase are influenced by the choice of the convection criterion. The Schwarzschild criterion predicts longer-lasting and faster-rotating Cepheids than the Ledoux criterion. The Ledoux criterion predicts the most massive Cepheids.
% RSG luminosity
\item Red supergiants computed with the Ledoux criterion reach a lower maximum luminosity than those computed with the Schwarzschild criterion.
% 20-40Msol
\item The impact of the criterion for convection on stars with initial masses between 20 and 40\,M$_\odot$ is complex because of the interplay with rotation and mass loss. For instance, the Ledoux criterion produces non-rotating WR stars for lower initial masses than the Schwarzschild criterion, but the inverse occurs for rotating stars.
% 60-120Msol
\item The impact of changing the convective criterion on stars whose initial masses are above 60\,M$_\odot$ is modest because their evolution is dominated by mass loss.
% Final models
\item The main differences relative to the endpoint of stellar evolution occur for stars with initial masses of 25 to 40\,M$_\odot$. For the 25\,M$_\odot$ the Ledoux criterion predicts the final compact objects to be black holes, while the Schwarzschild criterion predicts neutron stars. If these stars still undergo a supernova, the Ledoux criterion predicts much stronger helium signatures than the Schwarzschild criterion because the amount of helium in their envelope is larger.
% Clusters
\item Our population synthesis models imply different number ratios of evolved stars depending on the convection criterion, especially for non-rotating stars: The Schwarzschild criterion predicts much higher ratios of BSGs to RSGs and WCs to WNs than the Ledoux criterion. Conversely, the Ledoux criterion predicts a higher ratio of WRs to RSGs.
\end{itemize}
We recall, however, that the comparison presented in this work was made for a given choice of mass-loss rates, step overshoot, physics of rotation, and metallicity, and only for single-star evolution. Without performing detailed evolutionary calculations, it is not possible to know how the picture we presented in this work would change when one, a few, or all of the physical ingredients would change.

\begin{acknowledgements}
The authors would like to thank the anonymous referee for their very insightful comments which greatly improved the quality of this manuscript. The authors have received funding from the European Research Council (ERC) under the European Union's Horizon 2020 research and innovation programme (grant agreement No 833925, project STAREX).
\end{acknowledgements}

\bibliographystyle{aa}
\bibliography{References}

\end{document}